\documentclass[journal,12pt,onecolumn,draftclsnofoot]{IEEEtran}

\usepackage{amsmath}
\usepackage{cite}
\usepackage{comment}
\includecomment{fullversiononly} %
\excludecomment{notinfullversion} %

\usepackage{amssymb,amsthm}
\usepackage{amsfonts}
\usepackage{bm}                     %
\usepackage{algorithm}
\usepackage[noend]{algpseudocode}
\usepackage[flushleft]{threeparttable}

\usepackage{pgfplots}
\usepackage{enumerate}

\usepackage{todonotes}

\usepackage{url}

\DeclareMathOperator{\E}{\mathbb{E}}
\DeclareMathOperator*{\argmax}{arg\,max}

\newcommand{\mc}[1]{\mathcal{#1}}
\newcommand{\vv}[1]{\mathbf{#1}}
\newcommand{\cov}{\mathrm{Co}}

\newcommand{\rank}{\mathrm{rank}}
\newcommand{\col}{\mathrm{col}}
\newcommand{\bPsi}{\bm{\Psi}}
\newcommand{\bh}{h}

\newcommand{\bB}{\mathbf{B}}

\newcommand{\bG}{\mathbf{G}}
\newcommand{\bH}{\mathbf{H}}

\newcommand{\bU}{\mathbf{U}}
\newcommand{\bX}{\mathbf{X}}
\newcommand{\bY}{\mathbf{Y}}

\newtheorem{theorem}{Theorem}

\newtheorem{lemma}{Lemma}

\begin{document}
\title{Utility Maximization for Multihop Wireless Networks Employing BATS Codes}

\author{Yanyan~Dong,~Sheng~Jin,~Yanzuo~Chen,~Shenghao~Yang~and~Hoover~H.~F.~Yin
\thanks{This paper was presented in part at 2020 IEEE International Conference on Communications.}
\thanks{Y.~Dong, Y.~Chen and S.~Yang are with the School of Science and Engineering, The Chinese University of Hong Kong, Shenzhen, Shenzhen, China. S.~Jin is with Carnegie Mellon University. H.~Yin is with the n-hop technologies Limited, Hong~Kong, China and the Institute of Network Coding, The Chinese University of Hong Kong, Hong Kong, China. S.~Yang is also with Shenzhen Key Laboratory of IoT Intelligent Systems and Wireless Network Technology and Shenzhen Research Institute of Big Data, Shenzhen, China. Emails: \mbox{yanyandong@link.cuhk.edu.cn}, shengj@andrew.cmu.edu, yanzuochen@link.cuhk.edu.cn, shyang@cuhk.edu.cn, hfyin@inc.cuhk.edu.hk}
\thanks{This work was funded in part by the Shenzhen Science and Technology Innovation Committee (Grant JCYJ20180508162604311, ZDSYS20170725140921348).}
}
\maketitle

\begin{abstract}
  BATS (BATched Sparse) codes are a class of efficient random linear network coding
  variation that has been studied for multihop wireless networks
  mostly in scenarios of a single communication flow. Towards
  sophisticated multi-flow network communications, we formulate a
  network utility maximization (NUM) problem that jointly optimizes
  the BATS code parameters of all the flows and network
  scheduling. The NUM problem adopts a batch-wise packet loss model
  that can be obtained from the network local statistics without any
  constraints on packet loss patterns. Moreover, the NUM problem
  allows a different number of recoded packets to be transmitted for
  different batches in a flow, which is called adaptive recoding.  Due
  to both the probably nonconcave objective and the BATS code-related
  variables, the algorithms developed for the existing flow
  optimization problems cannot be applied directly to solve our NUM
  problem.  We introduce a two-step algorithm to solve our NUM
  problem, where the first step solves the problem with nonadaptive
  recoding schemes, and the second step optimizes adaptive
  recoding hop-by-hop from upstream to downstream in each flow.  We
  perform various numerical evaluations and simulations to verify the
  effectiveness and efficiency of the algorithm.
\end{abstract}

\section{Introduction}

Multihop wireless networks will play a crucial role in the future of
Internet of Things, where the communication from a source node to a
destination node may go through multiple intermediate network nodes
connected by wireless communication links.  Compared with the wireline
network design based on dedicated and reliable links, a more
sophisticated network design is required for wireless networks due to
the open, shared and dynamic wireless communication media.

\subsection{Background about Network Utility Maximization}

Wireline communication links, such as optical fiber and twisted cable, are highly reliable. Internet was built with wireline communication links in early days and has mechanisms designed based on the assumption of reliable links.  For example, intermediate nodes only perform \emph{store-and-forward}, i.e., only correctly received packets are forwarded to the next hop. Store-and-forward is optimal for communicating through networks formed by the concatenation of multiple reliable links. 
The network utility maximization framework originated from the analysis of Internet design naturally assumes reliable network links~\cite{kelly1998rate,low1999optimization}, where packet flow conservation is implied by store-and-forward.\footnote{Here packet flow conservation means that for a non-source, non-destination network node, the number of correctly decoded packets per unit time is the same as the number of outgoing packets per unit time.} %
Therefore, many network utility maximization problems for wireless networks also assume the wireless links to be highly reliable~\cite{lin2006tutorial,neill2008wnum}.

For wireless communications, however, link reliability cannot be guaranteed without sacrificing communication rate and latency. The wireless channel status is time-varying and cannot be known accurately due to interference, multipath fading, mobility, etc. The communication will fail with a high probability when the rate is higher than the instantaneous channel capacity, a phenomenon called \emph{outage}. Most modern wireless communication systems employ an Adaptive Modulation and Coding (AMC) mechanism to track the channel changes~\cite{goldsmith1998amc,qiu99amc}. Existing AMC algorithms try to maximize the transmission throughput (the transmission rate times the packet success rate) subject to a packet reliability constraint. %
In general, if we remove the reliability constraint, AMC can achieve a higher transmission throughput~\cite{weber2005transmission}. If an unlimited number of retransmissions are allowed for each packet, the wireless link can be reliable with an unbounded link latency. In practice, the number of retransmissions of a packet must be limited in order to bound the latency.

For networks without the link reliability assumption, the network utility maximization framework which assumes reliable link capacity can only provide a performance upper bound that cannot be achieved practically (to be further discussed in this paper). When each link has a certain packet loss rate and the packet loss events are independent, a NUM problem based on store-and-forward which is also called the leaky-pipe model has been studied in~\cite{gao2009cross}. In this model, it was shown that when each link has the same packet loss rate, the receiving rate at the destination node decreases exponentially as the number of hops increases.

\subsection{Motivation of This Paper}

Network coding provides a more general network communication framework than store-and-forward by allowing an intermediate network node to generate new packets~\cite{flow, linear, alg}.
It is well-known that \emph{random linear network coding (RLNC)} achieves the capacity of multihop
wireless networks with packet loss in a general setting
\cite{random, Dana2006,Lun2008}.
Network utility maximization has been studied for RLNC~\cite{wu06jsac,chen2007rc,khreishah2008rc,zhang2009mp,traskov12}. 
However, the classical RLNC approach is not
efficient for practical multihop wireless networks due to its high
computational and storage costs and also its high coefficient vector
overhead. To resolve the above issues, early implementations of RLNC partition the packets to be transmitted into small disjoint subsets known as \emph{batches}, \emph{generations}, \emph{chunks}, etc., and then apply RLNC for each batch separately \cite{chou03,gkan05}.
As the disjoint batches must be decoded individually, the end-to-end transfer matrix of each batch must be of full rank, which causes a scheduling issue for the disjoint batches \cite{maym06,bin_expander15}.

Compared with disjoint batches, it is more efficient to use batches with overlapped packets
\cite{Silva2009,Heidarzadeh2010,yaoli11, bin_expander15} and coded
batches~\cite{yang11ac,Mahdaviani12,Mahdaviani13,bin_ldpc16} as
the batches can help each other during decoding.  These low complexity RLNC schemes are
also collectively called \emph{batched network coding}. Among the
existing designs of batched network coding, BATS (BATched Sparse) codes have the
best overall performance in terms of computational complexity,
throughput and latency \cite{yang14bats,yang17monograph}.

A BATS code consists of an outer code and an inner code. The outer code is a
matrix generalization of a fountain code, which can generate a
potentially unlimited number of \emph{batches}. Each batch consists of
a certain number of coded packets, where this number is called the
\emph{batch size}. The inner code applies linear network coding,
which is also known as \emph{recoding}, at the intermediate network
nodes. Recoding is applied only to the packets belonging to the same
batch. All the batches in a BATS code can be decoded jointly.  It is
not required that all the batch transfer matrices are of full rank as for
RLNC with disjoint batches.  The achievable rate of BATS codes can
approach the expected rank of the \emph{batch transfer matrices},
which is also an upper bound on the achievable rate.

BATS codes have been extensively studied for multihop wireless networks \cite{huang14mobihoc,yang14a,xu2016two,yang18wuwnet,yin20entropy,zhou19}.
Towards sophisticated network communication protocols based on BATS codes,
\emph{network utility maximization (NUM)} problems for adaptive allocation of network resources, including congestion control and link scheduling, must be studied for BATS codes.
A packet flow generated by BATS codes has two
fundamental differences compared with a packet flow in a network employing store-and-forward:
\begin{enumerate}
\item There is no packet flow conservation at the intermediate network nodes because recoding generates new packets for the batches.
\item The end-to-end communication performance of a packet flow is not measured by the number of packets received per unit time. 
\end{enumerate}
Due to the different nature of packet flow when using BATS codes, the existing NUM models for store-and-forward and RLNC cannot be applied directly. A recent work~\cite{zhou20} studied the joint optimization of link scheduling and BATS code recoding in a single communication flow with independent packet loss. Here we study the NUM problem with multiple flows and with more general packet loss and recoding models.

\subsection{Our Contributions}

We have two main contributions in this paper: the NUM problem formulation and the algorithms for solving the NUM problem.

\subsubsection{NUM Problem Formulation}

We propose a general NUM %
problem where each communication flow formed by a path in the
network employs a BATS code for reliable end-to-end communications.
Each flow has a \emph{batch rate} measured by the number of batches
transmitted by the source node per unit time. As no new batch is
generated at the intermediate network nodes, the ``batch flow'' is
conservative.\footnote{Here batch flow conservation means that for a non-source, non-destination network node, the number of incoming batches per unit time is the same as the number of outgoing batches per unit time.} The variables of the NUM problem include the batch rate
of each flow, the recoding parameters for each flow at each network
node, and the scheduling rate vector.  Our NUM problem jointly
optimizes the total utilities of all the flows subject to certain
scheduling constraints. For each flow, the end-to-end performance is
measured as the product of the batch rate and the expected rank
of the batch transfer matrix, which is the maximum achievable
throughput of BATS codes.

Our NUM problem incorporates a general recoding framework called
\emph{adaptive recoding}
\cite{tang16schedu,yin_adaptive19,yin19overhearing,yin21intrablock,yin21impact,wang2021smallsample,xu2018,yin_adaptive16},
which allows a different number of recoded packets to be transmitted
for different batches in the same flow at a network node. In contrast, the recoding schemes which generate
the same number of recoded packets for all batches are called 
\emph{nonadaptive recoding}. It is clearly a waste of resources in nonadaptive recoding for transmitting some packets for the batches with rank-$0$ transfer matrices as they certainly contain no information. The recoding scheme studied in \cite{zhou19,zhou20} erases the batches with rank-$0$ transfer matrices and then uses the same number of recoded packets for all the remaining batches, and hence this scheme is a special case of adaptive recoding.
Existing works
have shown that the network throughput can be improved if more recoded
packets are generated for the batches having higher ranks in a single communication flow~\cite{xu2018,yin_adaptive16,yin19recoding}.
Our work further evaluates the performance of adaptive recoding in a network with multiple communication flows.

In the literature about batched network coding, the mostly used packet loss patterns are the independent packet loss and the Gilbert-Elliott packet loss model~\cite{gilbert1960capacity,elliott1963estimates}.
The conference version of this paper~\cite{dong20icc} only studied nonadaptive recoding under the independent packet loss model.
In this paper, we propose a batch-wise packet loss model to formulate the NUM problem so that the formulation is independent of particular packet loss patterns.
The batch-wise packet loss model can be practically obtained using the packet loss statistics at the two ends of a communication link. 
We justify the sufficiency of using this model in the problem formulation with adaptive recoding. 

The technical details regarding the problem formulation are organized as follows.
We first give a brief introduction of BATS codes in Section~\ref{sec:bats}.
Then, we discuss the network model and adaptive recoding in Section~\ref{sec:network}.
After that, we present the details of the NUM problem formulation in Section~\ref{sec:num}.

\subsubsection{Algorithms for Solving the NUM Problem}

Many traditional counterparts of our NUM problem have concave objectives and convex constraints.
 In contrast, our NUM problem has an objective function
that may not be concave and constraints that may not be convex in general.
Moreover, the adaptive
recoding parameters for different nodes and flows give a
large number of variables for the optimization. Due to the complexity of
our problem, the algorithms developed for the existing flow optimization
problems can be trapped in local maxima with a poor objective value according to our experiments.

We propose a two-step algorithm for solving our NUM problem. In the first step, the
NUM problem is solved with the restriction that recoding is
nonadaptive, i.e., the same number of recoded packets is generated for
all batches received by a node in the same flow.  With this
restriction, the number of variables of the NUM problem is greatly reduced: the recoding transition matrix at each node for each flow can be represented by a single value, which is the number of recoded packets to be generated.
In the second step, based on the solution 
obtained in the first step, we perform a sequence of hop-by-hop adaptive
recoding optimizations in an upstream-downstream order for each flow separately. Each hop-by-hop adaptive recoding optimization involves only local batch-wise packet loss statistics and can be solved efficiently by the
algorithms proposed in~\cite{yin_adaptive19}.

For simplicity, the number of recoded packets to be generated is also called the \emph{recoding number}.
In the first step, optimizing the recoding numbers is a new component of our NUM
problem compared with the traditional ones. Following the dual-based
approach for solving the existing NUM problems, the recoding number
optimization can be decomposed for each flow individually.
A special case of the recoding number optimization problem has been studied in~\cite{zhou19} with uniform random linear recoding and independent packet loss.
For our batch-wise packet loss model and a general recoding scheme, the technique in~\cite{zhou19} cannot be applied. We
illustrate that alternatively optimizing the recoding numbers of a flow
may not achieve a good performance.
We provide a local-search algorithm for optimizing all the recoding numbers of a flow jointly, which achieves satisfactory results according to our numerical evaluations. See the details in Section~\ref{sec:algorithm_u}. 
In the second step, we reallocate the recoding resources of nonadaptive recoding to gain the advantage of adaptive recoding. The number of recoded packets for the batches of low ranks are reduced and the released resources are used either to transmit more recoded packets for the batches of higher ranks or to increase the batch rate for transmitting more batches. %
See the  details in Section~\ref{sec:algorithm_ad}.

We conduct some numerical evaluations to compare the performance of different algorithms for solving our NUM problem (called (AP)). We also derive a corresponding traditional NUM problem (called (UP)), whose optimal value is an upper bound on the optimal value of (AP). For a feasible solution of (AP), we define a term called \emph{utility ratio} as the geometric average of the ratios of the feasible throughput of each flow over the one obtained by solving (UP). The higher the utility ratio is, the closer
the utility of the solution is to the upper bound.

We evaluate $11$ instances of wireless networks where some links are shared by two communication flows.
For both the independent and the Gilbert-Elliott packet loss models, the two-step algorithm achieves a good fairness for all the cases.
For independent packet loss, the two-step algorithm can achieve about $1.7\%$ to $2.75\%$ higher utility ratio compared with nonadaptive recoding. The relatively small gain of using adaptive recoding for independent packet loss is expected as the ranks of the batches are highly concentrated. %
For Gilbert-Elliott packet loss,
the utility ratio for the two-step algorithm is about $6\%$ to $10.5\%$ higher than the corresponding nonadaptive solution in the first step for eight of the instances.

We build a simulator using ns-3~\cite{Riley2010} to observe certain behaviors of the flow optimization solution in real network packet transmissions. We test both the independent packet loss and the Gilbert-Elliott packet loss models. To see the robustness of our approach, we solve the NUM problem with the empirical batch-wise loss model statistics and then substitute the output parameters into the simulator. We observe that the buffer sizes at all network nodes are stable, and the throughput is consistent with the one obtained by solving the NUM problem. See details of the simulations in Section~\ref{sec:sr}.

\section{BATS Code Basics}
\label{sec:bats}

In this section, we briefly introduce BATS codes.
We refer readers to \cite{yang17monograph} for a detailed discussion.
Fix a \emph{base field} of size $q$ and two positive integers $K$ and $T$.
Suppose the data to be transmitted consists of $K$ \emph{input packets}.
Each packet is regarded as a column vector of $T$ symbols in the base field.
We equate a set of packets to a matrix formed by juxtaposing the packets in this set.
A \emph{uniformly random matrix} is a matrix having uniform i.i.d. components over the base field.

\subsection{Encoding and Recoding}

The encoder of a BATS code generates the \emph{coded packets} in the \emph{batches}.
Let $M$ be a positive integer called the \emph{batch size}. For
$i=1,2,\ldots$, the $i$th batch, denoted by $\bX_i$, is generated from the input
packets using the following steps:
\begin{enumerate}
	\item Sample a \emph{degree distribution} $\bPsi=(\Psi_1, \ldots,\Psi_K)$ which returns a \emph{degree} $d_i$ with probability $\Psi_{d_i}$.
	\item Uniformly randomly choose $d_i$ input packets from all the input packets and juxtapose them into a matrix $\bB_i$.
	\item Form a $d_i\times M$ uniformly random matrix $\bG_i$ called the \emph{batch generator matrix}.
	\item The $i$th batch $\bX_i$ is generated by
  \begin{equation*}
    \bX_i = \bB_i \bG_i. %
  \end{equation*}
\end{enumerate}
We assume that the destination node knows the batch generator matrices.
This can be achieved by sharing the same seeds and the same pseudorandom number generator at the source node and the destination node.
\emph{Recoding} is a linear network coding scheme that is restricted to be applied to the packets belonging to the same batch.
The packets to be transmitted for a batch are the recoded packets of the batch generated by linear combinations of the received packets of the batch.
The number of recoded packets to be generated and the coefficients of the linear combinations depend on the recoding scheme.
We defer the discussion of the recoding schemes to Section~\ref{sec:adaptive} after introducing the network model.

\subsection{Batch Transfer Matrix}

As the recoding operation at each node is linear, the  transformation of each batch from
the source node to a network node that receives this batch is a linear operation.
Let $\bH_i$ be the \emph{batch transfer matrix} of the $i$th batch and $\bY_i$ be the
received packets of the $i$th batch at a network node. We have
\begin{equation}\label{eq:trans}
  \bY_i = \bX_i \bH_i = \bB_i\bG_i\bH_i.
\end{equation}
There are $M$ rows in $\bH_i$, where each row is formed by the coefficients of a coded packet in $\bX_i$.
The number of columns of $\bH_i$ corresponds to the number of packets of the $i$th batch received by the network node.
If no packet is
received for the batch, then both $\bY_i$ and $\bH_i$ are empty matrices.
More details about the batch transfer matrices will be discussed in
Section~\ref{sec:adaptive}. Here we introduce how the batch transfer
matrices can be known at a network node. Right after a batch is
generated, a coefficient vector is attached to each of the $M$
packets of the batch, where the coefficient vectors of all the packets of a
batch are linearly independent of each other. The recoding performed on the packets is also performed on the coefficient vectors.
One typical choice is that the juxtaposition of the coefficient vectors forms an
$M\times M$ identity matrix. By this choice, the transfer matrix of a batch can be recovered at each network node that receives this batch by juxtaposing the coefficient vectors of
the received packets of this batch.

The rank of the batch transfer matrix of a batch is also called
the \emph{rank of the batch}. Let $\bh = (\bh(r), r=0, 1, \ldots, M)$ be the underlying probability distribution of the ranks of the batch transfer matrices, which is also called the \emph{rank distribution}.
The empirical rank distribution of $\bH_i$, $i = 1, 2, \ldots$ converges almost surely towards $\bh$, which is a
sufficient statistics for designing the degree distribution.
The upper bound on the achievable rates of a BATS code is 
\begin{equation*}
	\E[\bh]\triangleq \sum_{i=1}^M i \bh(i),
\end{equation*}
which is also called the \emph{expected rank}. This upper bound can be achieved using random linear outer code \cite{yang10bf} with Gaussian elimination decoding. A major advantage of BATS codes is that this upper bound can be achieved by using much more efficient encoding and decoding methods.

\subsection{Decoding}

We can use the \emph{belief propagation (BP)} decoding algorithm to decode a BATS code efficiently.
Suppose that $n$ batches are received at the destination node.
Recall that the decoder of a BATS code knows the batch generator matrices $\bG_i$ and the batch transfer matrices $\bH_i$ of all the batches.
In other words, the decoder knows the linear systems of
equations in \eqref{eq:trans} for $i=1,\ldots,n$.
A batch with the generator matrix $\bG$ and the
batch transfer matrix $\bH$ is said to be \emph{decodable} if
$\rank(\bG\bH)$ is equal to its degree. The BP decoding includes
multiple iterations. In the first iteration, all the decodable batches
are decoded by solving the associated linear system of equations
in \eqref{eq:trans}, and the input packets involved in these decodable
batches are recovered. In each of the following iterations, undecoded
batches are first updated: for each undecoded batch, all the recovered input packets involved in the batch are substituted into the associated linear system and the degree
of the batch is reduced accordingly. Then, the batches which become
decodable after the update are decoded, and the input packets
involved in these decodable batches are recovered. The BP decoding
stops when there is no more decodable batch.

In the existing theory of BATS codes, a sufficient condition about the degree distribution was
obtained such that the BP decoding can recover a given fraction on the
number of input packets with high probability. A degree
distribution that satisfies the sufficient condition can be obtained
for a given rank distribution
$\bh$ by solving a linear programming problem~\cite{yang14bats}.
To have all the input packets solved by BP
decoding, precoding can be applied as in the Raptor codes.
This theory guarantees
that BATS codes can achieve a rate very close to $\E[\bh]$ with low
computational complexity.

When the number of packets for encoding is relatively small, BP decoding tends to stop before decoding a large
fraction of the input packets. Though we can continue decoding by Gaussian
elimination, the computational complexity is high. A better
approach is to use \emph{inactivation decoding}: when BP decoding
stops, an undecoded input packet is marked as inactive and
substituted into the batches as a decoded packet to resume the BP
decoding procedure.  Inactivation decoding reduces the complexity of
Gaussian elimination and improves the success probability of BP
decoding. See \cite{yang17monograph} for a detailed discussion of
inactivation decoding for BATS codes and the corresponding precoding design.

As a summary of this part, the end-to-end expected rank $\E[\bh]$ provides a precise performance measure of using BATS codes for network communication. 
 
\section{Network Communications Employing BATS Codes}
\label{sec:network}

In this section, we present the use of BATS codes for reliable end-to-end communications in a multihop wireless network model
similar to the one in \cite{lin2006tutorial}. %
In particular, we will discuss in detail how to employ \emph{adaptive recoding} at the network nodes.

\subsection{Network Model}

A multihop wireless network is modeled as a directed graph $\mc G(\mc V,\mc E)$, where $\mc V$ and $\mc E$ are sets of nodes and edges respectively.
A communication link in the network that transmits packets from node $u$ to node $v$ is modeled by an edge $(u,v)\in \mc E$.
Therefore, we also refer to an edge as a \emph{link}.
Each link $e = (u,v)$ is associated with a communication rate $c_e$ packets per unit time.
That is, node $u$ can transmit at most $c_e$ packets per unit time using the link $e$.

Due to interference and noise, not all the packets transmitted by node $u$ can be successfully received by node $v$.
Interference between different links is a characteristic of wireless networks due to the open wireless communication media.
For each link $e$, we have a set $\mc I_e$ of interfering links.
A \emph{collision} occurs when a link $(u,v)$ is scheduled to transmit with one or more interfering links simultaneously.
If a collision occurs, node $v$ cannot receive the packets transmitted by node $u$.
If no collision occurs, i.e., $(u,v)$ is scheduled to transmit without any interfering link in $\mc I_{(u,v)}$ transmitting simultaneously, then node $v$ may receive the packet transmitted by node $u$.
In the discussion of this paper, we assume that certain network scheduling mechanism is applied so that collision is avoided.
Without collision, it is also possible that a transmitted packet cannot be received correctly due to noise and signal fading.
The event that node $v$ does not receive the packet from $u$ is called \emph{packet loss}. A general packet loss model will be introduced in Section~\ref{sec:ana}.

We define a \emph{(link) schedule} by $s = (s_e, e\in \mc E)$ where $s_e = 1$ indicates that the link $e$ is scheduled to transmit and $s_e=0$ otherwise.
If $ s$ has no collision, i.e., $s_{e'}=0$ for all $e'\in \mc I_e$ if $s_e=1$ for any $e\in \mc E$, then the schedule is called \emph{feasible}.
$\mc S$ denotes the collection of all feasible schedules.
For a feasible schedule $s\in \mc S$, define a rate vector $ s'= (c_es_e, e\in \mc E)$, where $c_es_e$ is the maximum communication rate that link $e$ can support when the schedule $s$ is in used.
The \emph{rate region} is defined by
\begin{equation}\label{eq:rate}
  \mc R := \{(c_es_e, e\in \mc E)\mid s \in \mc S\}.
\end{equation}
The convex hull of $\mc R$ is denoted by $\cov(\mc R)$, which is the collection of all achievable scheduling rate vectors.

\subsection{Communication Flows Employing BATS Codes}
\label{sec:recoding}

For a directed edge $e=(u,v)$, we call $u$ the \emph{tail} of $e$ and $v$ the \emph{head} of $e$.
We say two edges $(u_1,v_1)$ and $(u_2,v_2)$ are \emph{consecutive} if $v_1=u_2$. 
A sequence of distinct links $\mc P=(e_{\ell} , \ell=1,\ldots,L)$ is called a \emph{flow (or path)} of length $L$ if $e_\ell$ and $e_{\ell+1}$ are consecutive for all $\ell=1,\ldots, L-1$.
A \emph{communication flow} is a flow in the network where the tail of the first edge is called the \emph{source node} and the head of the last edge is called the \emph{destination node}.

The network may have a finite number of communication flows concurrently.  Each
communication flow employs a BATS code for reliable end-to-end
communication from the source node to the destination node.  Fixed a
flow $\mc P =(e_\ell=(v_{\ell-1},v_\ell) , \ell=1,\ldots,L)$ of length $L$.
The source node $v_0$ uses the encoder of a BATS code to generate a
sequence of batches of batch size $M$. After that, each network node
in the flow performs recoding on the batches. In particular, the source
node $v_0$ performs recoding on the original $M$ packets of a batch,
and an intermediate network node $v_\ell$, $\ell = 1,\ldots,L-1$, performs
recoding on the received packets of a batch. %

The recoded packets can be generated using
different approaches as discussed in~\cite{yang17monograph}, e.g.,
\emph{uniformly random linear recoding} and \emph{systematic random linear recoding}. For completeness, we briefly introduce these approaches here.
Let $m$ and $r$ be two nonnegative integers.
Suppose $r$ linearly independent packets are received for a batch at node $u$ and denote by $\bY$ the matrix formed by these received packets.
That is, $\bY$ has $r$ columns.
Suppose node $u$ needs to transmit $m$ packets for this batch.
The transmitted packets are generated by $\bY\Phi$ where $\Phi$ is an $r\times m$ matrix called the \emph{recoding generator matrix}.
In literature, there are two mainstreams of random linear recoding.

\begin{itemize}
\item 
Uniformly Random Linear Recoding:
All the transmitted packets are generated by \emph{random linear combinations}, i.e., 
$\Phi$ is a uniformly random matrix over the base field (or a subfield of the base field). With a properly chosen $m$, this method asymptotically achieves the min-cut from the source node to the destination node when the batch size $M$ is sufficiently large.

\item
  Systematic Random Linear Recoding:
  This recoding approach transmits the linearly independent received packets before generating new packets by random linear combinations.   When $m\leq r$, $m$ received packets with linearly independent coefficient vectors are transmitted. When $m>r$, the $r$ linearly independent received packets and $m-r$  packets generated by random linear combinations are transmitted. With certain row and column permutations, $\Phi$ is in the form of $
\begin{bmatrix}
  \mathbf I_r & \bU
\end{bmatrix}$, where $\mathbf I_r$ is an $r\times r$ identity matrix and $\bU$ is a uniformly random matrix.   
\end{itemize}
When the base field is sufficiently large, these two approaches have almost the same performance. Uniformly random linear recoding is easier to analyze in many cases, so we use this recoding approach without otherwise specified.

\subsection{Adaptive Recoding}
\label{sec:adaptive}

We introduce a general adaptive recoding
framework~\cite{yin_adaptive19} for a length-$L$ flow
$\mc P =(e_i=(v_{i-1},v_i) , i=1,\ldots,L)$. Recall that each node can check the coefficient vector attached
to each packet it has received and hence it knows the rank of the
batches it has received. The maximum rank of a batch is the batch size
$M$. At the source node, each batch generated by the encoder has a
rank $M$. For adaptive recoding, the number of recoded packets to be
transmitted for a received batch is a random variable that depends on
the rank of this batch.

For an integer $m\geq 0$, the source node transmits $m$ recoded packets of the batch with
probability $p_{e_1}(m|M)$. 
The recoding at the intermediate network nodes can be formulated
inductively.  At an intermediate network node $v_i$, $i = 1,\ldots,L-1$, for an integer $m\geq 0$, the node transmits $m$ recoded packets for a received batch of rank $r$ with probability $p_{e_{i+1}}(m|r)$. %
We denote $p_{e_i}$ as a stochastic matrix where its $(r,m)$ entry is $p_{e_i}(m|r)$.
Denote by $\bh_v=(\bh_{v}(r), r = 0,1,\ldots,M)$ the rank distribution of a batch received at
node $v$. At the source
node $v_0$ of flow $\mc P$, we have $\bh_{v_0}=(0,\ldots,0,1)$. In practice, $\bh_v$ can be obtained empirically at node $v$ by using the coefficient vectors attached to the received packets.
In general, $\bh_{v_0}, \ldots, \bh_{v_L}$ form a Markov chain for the
batches transmitted in flow $\mc P$.

Fix an edge $e=(u,v)\in \mc P$. The expected number of transmitted packets per batch on edge $e$ is
\begin{equation}\label{eq:mee}
  \overline{m}_{e} := \sum_{r}\sum_m m p_{e}(m|r) \bh_{u}(r).
\end{equation}
\begin{notinfullversion}
The $(r,m)$ entry of the gradient of $ \overline{m}_{e} $ with respect to $p_e$ is
\begin{equation}
	\frac{\partial  \overline{m}_{e} }{\partial p_e(m|r)}=mh_{u,r}, \label{eq:gradient1}
\end{equation}
i.e., $\frac{\partial  \overline{m}_{e} }{\partial p_e}=(mh_{u,r})_{r,m}$, which will be used in our algorithm.
\end{notinfullversion}
When $\overline{m}_e$ has an upper bound $m_e$, a local optimization problem has been studied in~\cite{tang16schedu,yin_adaptive19} to maximize the expected rank at node $v$, which can be written as
\begin{equation} \label{eq:aropt}
  \begin{IEEEeqnarraybox*}[][c]{rCl}
  \max_{p_e} & \quad & \E[\bh_v] \\
  \text{s.t.} & & \overline{m}_e \leq m_e,
\end{IEEEeqnarraybox*}
\end{equation}
where the explicit formula of $ \E[\bh_v] $ will be discussed later on.
We say that $p_e$ is \emph{almost deterministic} if there exists $t_e=(t_e(r),r=0,1,\ldots,M)$ such that
\begin{equation*}
  p_e(m|r) =
  \begin{cases}
	  t_e(r) - \lfloor t_e(r) \rfloor & \text{if } m = \lfloor t_e(r) \rfloor + 1, \\
	  1 - (t_e(r) - \lfloor t_e(r) \rfloor) & \text{if } m = \lfloor t_e(r) \rfloor, \\
    0 & \text{otherwise}.
  \end{cases}
\end{equation*}
It has been shown in~\cite{yin_adaptive19} that under certain technical concavity condition, there exists an almost deterministic optimizer for \eqref{eq:aropt}. The concavity condition can be satisfied when the packet loss pattern is a stationary stochastic process~\cite{yin_adaptive19}.

Two special cases of adaptive recoding are also of our interest:
\begin{itemize}
\item 
Before adaptive recoding was proposed, for BATS codes, the same number of recoded packets is transmitted for all the batches in a flow, i.e., $p_{e_i}(m_{e_i}|r) = 1$ for a certain integer $m_{e_i}$ for all $r$.
We refer to this kind of recoding scheme as \emph{nonadaptive recoding}. Adaptive recoding can achieve a higher expected rank than nonadaptive recoding under the same average number of recoded packets constraint \cite{yang14a,tang16schedu,yin_adaptive16}. 

\item When the batch size is $M=1$, store-and-forward performed in many existing network communication protocols can be regarded as a special recoding scheme, where $p_{e_i}(0|0)=1$ and $p_{e_i}(1|1)=1$ for $i=1,\ldots,L$. In other words, each node can only transmit the packets it has received. 
\end{itemize}

\subsection{Batch-wise Packet Loss Model}
\label{sec:ana}

We consider a \emph{batch-wise packet loss model} in this
paper: for each edge $e=(u,v)$, when $m$ packets of a batch are
transmitted by node $u$, node $v$ receives $r$ packets of the batch
with probability $q_e(r|m)$. We suppose the stochastic matrix $q_e=(q_e(r|m))$ is
fixed for each edge $e$. This is an artificial packet loss model that
includes independent packet loss as a special case.
For independent packet loss, we have
\begin{equation}\label{eq:il}
  q_e(r|m) = \binom{m}{r} (1-\epsilon_e)^r \epsilon_e^{m-r},
\end{equation}
where $\epsilon_e$ is the packet loss rate on edge $e$.
Practically, $q_e(r|m)$ can be obtained locally at node $v$
by counting the number of received packets for each batch.  The
advantage of this model is that it keeps the analysis of BATS code
feasible without knowing the complete packet loss statistics.

Consider a flow $\mc P =(e_i=(v_{i-1},v_i) , i=1,\ldots,L)$
employing the adaptive recoding scheme introduced above.  The
 distribution $\bh_v$ for any node $v$ in the flow can be
derived analytically for the batch-wise independent packet loss model.
Denote by $P_{v_l}$ the $(M+1)\times (M+1)$ transition matrix from
$\bh_{v_{l-1}}$ to $\bh_{v_l}$. For example, when applying uniformly random linear recoding to generate recoded packets, as derived in
\cite[chap.~4]{yang17monograph}, the $(i,j)$ entry of $P_{v_l}$ is
\begin{equation*}
	P_{v_l}[i,j] = \begin{cases}
		\sum_{m\geq j}p_{e_l}(m|i)\sum_{k=j}^{m}q_{e_l}(k|m) \zeta^{i,k}_{j} & \text{if } j \leq i,\\
		0 & \text{otherwise},
	\end{cases}
\end{equation*}
where $\zeta^{i,k}_{j}$ is the probability that an $i\times k$ uniformly random matrix over the base field has rank $j$ which has closed-form expressions and therefore can easily be computed. %
Hence, we have $\E[\bh_{v_L}] = \bh_{v_L} (0,1,...,M)^T$ where
\begin{equation*}
  \bh_{v_L} = \bh_{v_0} P_{v_1}\cdots P_{v_l} \cdots P_{v_L}.
\end{equation*}

For systematic random linear recoding, we need a further assumption to derive the transition matrix: All the packets of a batch are transmitted subject to a permutation chosen uniformly at random. With this assumption, when $k$ packets are received for $m$ transmitted packets, all the combinations of $k$ packets among the $m$ packets are received with the same probability.
Hence for systematic random linear recoding, a transition matrix can also be derived similarly as~\cite[chap.~4]{yang17monograph} for the batch-wise packet loss model. 

More discussions about the batch-wise packet loss model are given in Appendix~\ref{app:lossmodel}. These discussions provide some theoretical guidances about using this model. For example, we will give a sufficient condition of the batch-wise packet loss model so that an almost deterministic optimizer exists for \eqref{eq:aropt}, which extends the corresponding result in~\cite{yin_adaptive19}.

\section{Network Utility Maximization Problems}
\label{sec:num}

In this section, we formulate a utility optimization problem for
multihop wireless networks with BATS codes and discuss some special cases of the problem.

\subsection{General Problem}

Consider a fixed number of communication flows in the network
where the $i$th flow is denoted by $\mc P^i$. Each flow employs
a BATS code independently. The source node of the flow $\mc P^i$
generates batches of batch size $M^i$ and transmits $\alpha_i$ batches per unit time, where $\alpha_i$ is also called the \emph{batch rate}.
The network nodes in $\mc P^i$ employ certain recoding schemes on the batches as we have described in Section~\ref{sec:recoding}.

For an edge $e=(u,v) \in \mc P^i$, we denote by $p_e^i(m|r)$ the probability of transmitting $m$ recoded packets on link $e$ for a batch of rank $r$ at node $u$ in flow $\mc P^i$.
Let $\bh^i$ be the rank distribution of a batch at the destination node 
in the $i$th flow.  The \emph{throughput} of the $i$th flow is
$\alpha_i\E[\bh^i]$, which can be achieved as we have discussed in Section~\ref{sec:bats}. 
We want to maximize the \emph{total utility} of all flows defined as
$\sum_i U_i(\alpha_i\E[\bh^i])$ where each $U_i$ is a certain non-decreasing concave utility function.

The constraints of the above utility maximization involve both the link capacities and link scheduling. For a batch in flow $\mc P^i$, by \eqref{eq:mee}, the average number of packets transmitted on edge $e=(u,v)$ is
\begin{equation*}
  \overline{m}_e^i = \sum_r\sum_m m p_e^i(m|r)h_u^i(r),
\end{equation*}
where $h_u^i$ is the rank distribution of a batch at node $u$ in flow $\mc P^i$.
In practice, $\overline{m}_e^i$ can be counted at each network node locally.
For each scheduling rate vector $s = (s_e,e\in \mc E) \in \cov(\mc R)$, the average  number of packets transmitted on edge $e$ per unit time should be no more than $s_e$, i.e.,
$\sum_{i\colon e\in \mc P^i} \alpha_i \overline{m}_e^i \leq s_e$.

As a summary of the above problem formulation, consider the communication flows $ (\mc P^i)_i $ in a network with the edge set $ \mc E $ and the scheduling rate region $\mc R$. Each flow $\mc P^i$ has the batch rate $\alpha_i$ and recoding parameters in each link $  (p_e^i, e\in \mc P^i)$ as the variables for flow control. The variables of all flows are collectively written as $(\alpha_i,  (p_e^i, e\in \mc P^i))_i$. The utility maximization problem stated above can be written as follows:
\begin{equation} \tag{AP}
  \label{eq:1}
\begin{IEEEeqnarraybox*}[][c]{rCl}
  \max_{(\alpha_i, (p_e^i, e\in \mc P^i))_i,  s} & \quad & \sum_i U_i(\alpha_i\E[\bh^i]), \\
  \text{s.t.} & & \sum_{i\colon e\in \mc P^i} \alpha_i \overline{m}_e^i \leq s_e, \quad \forall e\in \mc E \\
  & & s = (s_e,e\in \mc E) \in \cov(\mc R).
\end{IEEEeqnarraybox*}
\end{equation}
Henceforth, we refer the above optimization problem as \eqref{eq:1}, where AP is the acronym of Adaptive Problem.

\begin{notinfullversion}
For the stochastic matrix $p_e^i$ in the above optimization problem,
we assume $p_e^i(0|0)=1$ and we only consider $p_e^i(m|r)$ for $0\leq m\leq M_0$
and $1\leq r\leq M$, where $M_0$ is the maximum number of
recoded packets that can be transmitted, e.g., $10 M$. Practically, it is
not feasible to transmit too many packets for a batch due to the
effect of latency. Analytically, due to the rate
constraints on each edge, 
increasing $M_0$ beyond a certain value has
negligible help for improving
the expected rank at the destination node.
\end{notinfullversion}

\subsection{Nonadaptive Recoding Problem}

Consider nonadaptive recoding with $p_e^i(m_e^i|r) = 1$ for a certain integer $m_{e}^i$ for all $r$.
The general adaptive recoding NUM problem \eqref{eq:1} becomes
\begin{equation}\tag{NAP}
  \label{eq:ur}
\begin{IEEEeqnarraybox*}[][c]{rCl}
  \max_{(\alpha_i, (m_e^i,e\in \mc P^i))_i, s} & \quad & \sum_i U_i(\alpha_i\E[\bh^i]), \\
  \text{s.t.} & & \sum_{i\colon e\in \mc P^i} \alpha_i m_e^i \leq s_e, \quad \forall e\in \mc E \\
  & & s=(s_e, e\in \mc E) \in \cov(\mc R).
\end{IEEEeqnarraybox*}
\end{equation}
We refer the above optimization problem as \eqref{eq:ur}, where NAP is the abbreviation of NonAdaptive Problem.

As a variation, we can let $f_e^i=\alpha_im_e^i$ to linearize the constraints and use
$\lfloor f_e^i/\alpha_i\rfloor$ to replace $m_e^i$ in $\bh^i$ to
remove the discrete variables. But the expected rank $\E[\bh^i]$ is not
necessarily concave nor continuous in terms of $f_e^i$ and $\alpha_i$.
These facts make the nonadaptive recoding NUM problem \eqref{eq:ur} usually more
difficult to solve than the traditional counterparts which we will discuss later.

For BATS codes, couple special cases of optimization problem \eqref{eq:ur} with only one flow have been studied in the literature.  When there is only one flow $\mc P$, \eqref{eq:ur} becomes
\begin{equation}
  \label{eq:2}
\begin{IEEEeqnarraybox*}[][c]{rCl}
  \max_{\alpha, (m_e,e\in \mc P), s} & \quad & \alpha\E [\bh] \\
  \text{s.t.} & & \alpha m_e \leq s_e, \quad \forall e\in \mc P \\
  & & s=(s_e,e\in \mc P) \in \cov(\mc R),
\end{IEEEeqnarraybox*}
\end{equation}
where $ \alpha $ is the batch rate of this flow, $(m_e, e\in \mc P)$ are the recoding numbers of each link, and $ h $ is the rank distribution of the batches received at the destination node in $ \mc P. $
Note that we omit the utility function as it is non-decreasing. 

\subsubsection{Single flow, no collision}
The recoding optimizations in \cite{yang17monograph} focus on the case that links have the same rate $c$ and no collision occurs so that $\cov(\mc R)=\{(s_e,e\in \mc E)\mid s_e \leq c\}$. This rate region is suitable for multi-ratio multi-channel wireless networks \cite{kyasanur2006multichannel}.
For $e\in \mc P$, suppose $m_{e'}$ is fixed for all $e'\neq e \in \mc P$, then $\E[h]$ is a non-decreasing function of $m_e$ for uniformly random linear recoding (see Appendix~\ref{app:erf} for the justification).
The optimization problem \eqref{eq:2} is equivalent to 
\begin{equation}
\label{eq:2+}
\begin{IEEEeqnarraybox*}[][c]{rCl}
\max_{\alpha, (m_e=m,e\in \mc P), s} & \quad & \alpha\E [\bh] \\
\text{s.t.} & & \alpha m = c, \quad \forall e\in \mc P ,
\end{IEEEeqnarraybox*}
\end{equation}
which can be solved by
\begin{align}
\label{eq:2-}
  \max_{ (m_e=m,e\in \mc P)} \frac{c\E[h]}{m},
\end{align}
 and  $ \alpha^* = c/m^* $ with $ m^* $ being the optimizer of \eqref{eq:2-}. Note that \eqref{eq:2-} can be solved easily by exploring a range of integer values of $m$.
\subsubsection{Single flow, all collision}
The collision model in \cite{zhou19} is that, only one link can transmit at a time, and the capacity of each link is $c$.
Hence  $\cov(\mc R)=\{(s_e,e\in \mc E) \mid \sum_{e}s_e \leq c\}$, where
this rate region is suitable for the case that all nodes are very close to each other. 
With this rate region, the optimization problem \eqref{eq:2} becomes
\begin{equation*}
\begin{IEEEeqnarraybox*}[][c]{rCl}
\max_{\alpha,  (m_e,e\in \mc P), s} & \quad & \alpha\E [\bh] \\
\text{s.t.} & & \alpha \sum_e m_e \leq c, \quad \forall e\in \mc P ,
\end{IEEEeqnarraybox*}
\end{equation*}
which can be solved by
\begin{equation}\label{eq:sub}
  \max_{  (m_e,e\in \mc P)} \frac{c\E[h]}{\textstyle{\sum_{e}m_e}},
\end{equation}
and $ \alpha^* = \frac{c}{\sum_e m_e^*} $ with $ (m_e^*, e\in \mc P) $ being the optimizer of \eqref{eq:sub}.

In \cite{zhou19,zhou20}, when no packet is received for a batch, no packet is transmitted for the batch; and when a positive number of packets are received, the same number of recoded packets are generated. This recoding can be regarded as a special adaptive recoding approach.
When independent packet loss is assumed, the probability that no packet is received for a batch is very low when the batch size is not too small, e.g., 16. Therefore, this modification of the problem has little effect on the objective and hence has a very similar performance as nonadaptive recoding.

\subsection{Traditional Cases: Store-and-Forward Recoding}

When the batch size $M^i=1$ for all flows, 
\eqref{eq:1} includes some classical network utility maximization problems as special cases when applying store-and-forward as discussed in Section~\ref{sec:adaptive}. 
Many existing works do not consider packet loss when there is no collision \cite{lin2006tutorial,shakkottai2008network}. In this case, the destination node of flow $ \mc P^i $ receives  $ \alpha_i $ packets per unit time and the optimization problem \eqref{eq:1} becomes
\begin{equation}
  \label{eq:to}
  \begin{IEEEeqnarraybox*}[][c]{rCl}
    \max_{(\alpha_i)_i,s} & \quad & \sum_i U_i(\alpha_i) \\
    \text{s.t.} & & \sum_{i\colon e\in \mc P^i} \alpha_i \leq s_e, \quad \forall e\in \mc E \\
    & & s=(s_e, e\in \mc E) \in \cov(\mc R).
  \end{IEEEeqnarraybox*}
\end{equation}

Suppose each link has a certain packet loss rate and the packet losses are independent, the optimization problem \eqref{eq:1} becomes the one of the leaky-pipe flow model \cite{gao2009cross}.  Consider a flow of $L$
links of identical rate, where the packet loss rate of each link is $\epsilon\in
(0,1)$. In the leaky-pipe model, the data rate decreases to
$(1-\epsilon)$ fraction hop-by-hop so that the receiving rate is only
$(1-\epsilon)^L$ of the rate of the source node, i.e., the end-to-end
throughput decreases exponentially when the number of hops increases.

\subsection{Cut-set Upper Bound}

To assist the evaluation of the optimality of solving \eqref{eq:1}, we provide an upper bound that corresponds to the cut-set bound in a multihop network under the condition that each link has a constant loss rate.
Consider a length-$L$ flow $\mc P=(e_\ell=(v_{\ell-1},v_\ell) , \ell=1,\ldots,L)$, where a BATS code of batch size $M$ is applied. 
Denote by $Y_\ell$ the random variable of the number of received packets for a batch at node $v_\ell$, where $\ell=1,\ldots,L$. 
Let $ h $ be  the rank distribution of the batches received  at node $ v_L $,
we have
\begin{IEEEeqnarray*}{rCl}
  \E[h]
  & \leq &  \E(\min\{M, Y_1,\ldots,Y_L\}) \\
  & \leq & \min\{M, \E(Y_1), \ldots, \E(Y_L)\}  \IEEEyesnumber \label{eq:up}
\end{IEEEeqnarray*}
which is the cut-set upper bound of the expected rank. 

Recall that we assume a certain network scheduling mechanism is applied so that collision is avoided.
Suppose each link $e=(u,v)$ has a loss rate $\epsilon_e$. Then $1-\epsilon_e$  is the ratio of the expected receiving rate at $v$ and the transmitting rate at $u$ on the link. Using the notations in \eqref{eq:1}, if $e\in \mc P^i$, then $(1-\epsilon_e) \overline{m}_e^i$ is the expected number of received packets for a batch at node $v$ in flow $\mc P^i$. 
By the upper bound shown in \eqref{eq:up}, we have
\begin{IEEEeqnarray*}{rCl}
  \E[h^i] & \leq & \min\{M^i,\overline{m}_e^i(1-\epsilon_e) , e\in\mc P^i\} \\
  & \leq & \min\{\overline{m}_e^i(1-\epsilon_e) , e\in\mc P^i\},
\end{IEEEeqnarray*}
where $ h^i $ is the rank distribution of the batches received at the destination node in flow $\mc P^i$.
By letting $f_e^i=\alpha_i\overline{m}_e^i(1-\epsilon_e)$ and replacing $\E[h^i]$ by the above upper bound, the optimization problem \eqref{eq:1} becomes
\begin{equation}\tag{UP}
  \label{eq:3}
\begin{IEEEeqnarraybox*}[][c]{rCl}
  \max_{(f^i_e,e\in \mc P^i)_i,s} & \quad & \sum_i U_i(\min\{f_e^i , e\in\mc P^i\}) \\
  \text{s.t.} & &  \sum_{i \colon e\in \mc P^i}f_e^i \leq s_e(1-\epsilon_e), \quad \forall e\in \mc E \\
  & & s=(s_e,e\in \mc E) \in \cov(\mc R).
\end{IEEEeqnarraybox*}
\end{equation}
We refer the above optimization problem as \eqref{eq:3}, where UP is the acronym of Upper-bound Problem.

The optimal value of \eqref{eq:3} is an upper bound on the optimal value of \eqref{eq:1}. This upper bound is also called the cut-set bound of the network communication capacity, and can be achieved when the batch size is unbounded.
This upper bound, however, is not achievable in practical scenarios where a limited batch size is required for delay and buffer size consideration. Optimization \eqref{eq:3} is of the form of \eqref{eq:to}, and hence can be solved by using, e.g., the algorithms introduced in \cite{lin2006tutorial,shakkottai2008network}.

\section{Algorithms for Nonadaptive Recoding Problem}
\label{sec:algorithm_u}

In this section, we discuss how to solve the nonadaptive recoding NUM problem \eqref{eq:ur}, which is a special case of the general adaptive recoding NUM problem~\eqref{eq:1}. This special case keeps the main features of the general problem and hence is a good starting point for studying the optimization algorithms.
If the recoding parameters $m_e^i$ are all fixed, problem~\eqref{eq:ur} becomes a weighted version of the traditional case \eqref{eq:3} and hence can be solved using the existing algorithms.  

The traditional network utility maximization problems of the form \eqref{eq:to} can be solved by primal, dual or primal-dual approaches~\cite{shakkottai2008network}. As the problem decomposition induced by the dual approach has the advantage of simplifying the complexity~\cite{lin2006tutorial}, we adopt this approach for our optimization problem. Our main purpose here is to illustrate how to handle the recoding parameters $m_e^i$. The guidelines provided here can be applied to the primal and the primal-dual approaches as well.

\subsection{Dual-Based Algorithm}

Associating a Lagrange multiplier $q_e$ for each inequality constraint in \eqref{eq:ur}, the Lagrangian is
\begin{IEEEeqnarray*}{Cl}
  & \sum_i U_i(\alpha_i\E[h^i]) - \sum_{e\in \mc E} q_e \left(\sum_{i\colon e\in \mc P^i} \alpha_i m_e^i - s_e\right) \\
  = & \sum_i \left[ U_i(\alpha_i\E[h^i]) - \alpha_i \sum_{e\in \mc P^i} q_e  m_e^i \right]  + \sum_{e\in \mc E} q_e s_e,
\end{IEEEeqnarray*}
where $q_e$ is an implicit cost for the link $e$. %
The dual problem of \eqref{eq:ur} is to minimize
\begin{IEEEeqnarray*}{rCl}
  \sum_i \max_{\alpha_i, m^i} \left[U_i(\alpha_i\E[h^i])-\alpha_i \sum_{e\in \mc P^i} q_e  m_e^i \right] + \max_{s \in \cov(\mc R)}\sum_{e\in \mc E} q_e s_e,
\end{IEEEeqnarray*}
where $m^i = (m_e^i, e\in \mc P^i)$.

Similar to \cite{lin2006tutorial}, we have the following solution of the dual problem. We use $t = 1,2,\ldots$ to denote the $t$th iteration of the algorithm.
Let $\alpha_i(t)$, $m_e^i(t)$, etc., be the values of the corresponding variables $\alpha_i$, $m_e^i$, etc., in the $t$th iteration. 
The batch rate $\alpha_i(t)$ and the recoding number $m^i_e(t)$ are updated by solving
  \begin{equation}\label{eq:update_recoding}
	  \mkern-6mu (\alpha_i(t), m^i(t))=\argmax_{\alpha_i, m^i} \left[U_i(\alpha_i\E[h^i])-\alpha_i \sum_{e\in \mc P^i} q_e(t)  m_e^i \right]
  \end{equation}
 for each flow $i$.
 The scheduling rate vector is determined by
  \begin{equation}\label{eq:update_scheduling}
	   s(t) = \argmax_{s\in \cov(\mc R)}\sum_{e\in \mc E} q_e(t) s_e.
  \end{equation}
 The Lagrange multipliers are updated by
  \begin{equation}\label{eq:update_mul}
	  \mkern-6mu q_e(t+1) = \left[q_e(t) + \gamma_t \left( \sum_{i\colon e\in \mc P^i}\alpha_i(t)m_e^i(t) - s_e(t) \right) \right]^+
  \end{equation}
  where $[x]^+ = \max\{0,x\}$ for real $x$, and $\gamma_t$, $t=1,2,\ldots$ is a sequence of positive step sizes such that the subgradient search converges, e.g., $ \sum_t \gamma_t = \infty $ and $ \sum_t \gamma_t^2 < \infty $.

  In the above algorithm, the optimization of the batch rates and recoding numbers for each flow is decoupled, so that the complexity is linear with the number of flows. Moreover, the scheduling update in \eqref{eq:update_scheduling} is the same as the one in the traditional problem (see \cite[eq.~(12)]{lin2006tutorial}), so that the same imperfect scheduling policies discussed in \cite{lin2006tutorial} can be applied to simplify the complexity of finding an optimal scheduling strategy. 
However, our problem is more complicated than the traditional ones in the following two aspects.

First, the subproblem \eqref{eq:update_recoding} optimizes the recoding numbers along a path jointly. As the  integer recoding numbers get involved in the complicated formula of $\E[h^i]$, experiments show that it is not optimal to alternately optimizes one recoding number while fixing the others. The exhaustive search of the optimal recoding numbers has exponential complexity in terms of the path length. To reduce the complexity, we limit each recoding number within a small neighboring range in each iteration to find a local optimum. We will elaborate how to solve \eqref{eq:update_recoding}  in Section~\ref{sec:rnu}.
  
Second, as the primal \eqref{eq:ur} is nonconcave, the optimizer of the dual problem may not be feasible for the primal one. Therefore, after obtaining a dual solution $\{(\tilde \alpha_i, (\tilde{ m}_e^i,e\in \mc P^i))_i\}$ by multiple rounds of updates using \eqref{eq:update_recoding}-\eqref{eq:update_mul},  we find a feasible primal solution by solving 
  \begin{equation}\label{eq:4}
    \begin{IEEEeqnarraybox*}[][c]{rCl}
      \max_{(\alpha_i)_i, s} & \quad & \sum_i U_i(\alpha_i\E[h^i]) \\
      \text{s.t.} & & \sum_{i\colon e\in \mc P^i} \alpha_i {\tilde{m}_e^i} \leq s_e, \quad \forall e\in \mc E \\
      & & s=(s_e, e\in \mc E) \in \cov(\mc R),
    \end{IEEEeqnarraybox*}
  \end{equation}
  with fixed recoding numbers $\tilde{m}_e^i$ of the dual solution so that $\E[h^i]$ is also fixed. The optimization problem \eqref{eq:4} is of the form of \eqref{eq:to}, and hence can be solved by using, say, the algorithms introduced in \cite{lin2006tutorial,shakkottai2008network}.

\subsection{The Updates in Each Flow}
\label{sec:rnu}

In this subsection, we discuss how to update the batch rate and recoding number in \eqref{eq:update_recoding}, i.e., the way to solve
\begin{equation*} %
  \max_{m^i} \max_{\alpha_i} \left[U_i(\alpha_i\E[h^i])-\alpha_i \sum_{e\in \mc P^i} q_e  m_e^i \right],
\end{equation*}
where $q_e\geq 0$.
Assume $\sum_{e\in \mc P^i} q_e>0$ since otherwise the problem is trivial.
As $U_i$ is concave, the inner maximization problem has a unique optimal solution $\alpha_i^*$ for any given $m^i$. Therefore, the essential problem here is to solve
\begin{equation}\label{eq:mm}
  \max_{m^i} \left[U_i(\alpha_i^*\E[h^i])-\alpha_i^* \sum_{e\in \mc P^i} q_e  m_e^i \right].
\end{equation}
When the utility function is the natural logarithm function, we have
$\alpha_i^* = \frac{1}{\sum_{e\in \mc P^i} q_e  m_e^i}$.
Hence, the $m^i$ solving
\begin{equation}
  \label{eq:mmlog}
  \max_{m^i} \frac{\E[h^i]}{\sum_{e\in \mc P^i} q_e  m_e^i}
\end{equation}
can also solve \eqref{eq:mm}.

The optimization problem \eqref{eq:mmlog} has been studied in~\cite{zhou19} for the case with independent packet loss and uniformly random linear recoding. For this special case, the method in~\cite{zhou19} is to relax $m_e^i$ to be real numbers and then solve the relaxed problem using certain existing continuous optimization solver to obtain the optimizer $(\tilde m_e^i,e\in \mc P^i)$.
After that, an exhaustive search is performed in the set $\{(m_e, e\in\mc P^i)\mid m_e\in \mathbb{N} ,\lfloor \tilde m_e^i \rfloor \leq m_e \leq \lceil \tilde m_e^i \rceil\}$ for the final solution, where $\mathbb{N}$ is the set of natural numbers.

The approach in~\cite{zhou19} discussed above depends on the special property of independent packet loss and hence cannot be extended for a general packet loss model. Here, we propose a local search algorithm for solving \eqref{eq:mm} and \eqref{eq:mmlog} with a general batch-wise packet loss model. For a vector $v =(v_e,e\in \mc P^i)$, define
\begin{equation*}
  \mc N(v) = \{(m_e,e\in \mc P^i) \mid m_e\in \mathbb{N},v_e-1 \leq m_e \leq v_e +1\}.
\end{equation*}
The algorithm starts with an initial vector $m(0) = (m_e(0), e\in \mc P^i)$ where $m_e(0)\geq 0$. In the $t$th iteration, $t=1,2,\ldots,$ the local search algorithm performs an exhaustive search in the set $\mc N(m(t-1))$ to find $m(t)$ that maximizes \eqref{eq:mm} (or \eqref{eq:mmlog} for natural logarithm utilities). The algorithm stops when $m(t)=m(t-1)$, which indicates that $m(t)$ is a local optimal solution in $\mc N(m(t))$.
For an edge $e$ such that $q_e>0$, as $\E[h^i]$ is upper bounded by the batch size, the optimal $m_e^i$ is also bounded.
When $q_e = 0$, the objective function of \eqref{eq:mm} is non-decreasing in $m_e^i$, and hence the above terminating condition may not be satisfied for any finite $t$.  We add an additional terminating condition that the algorithm stops when the improvement of the objective in an iteration is smaller than a certain threshold. As the objective is upper bounded and is increasing in each iteration, the additional terminating condition can guarantee the termination of the algorithm.

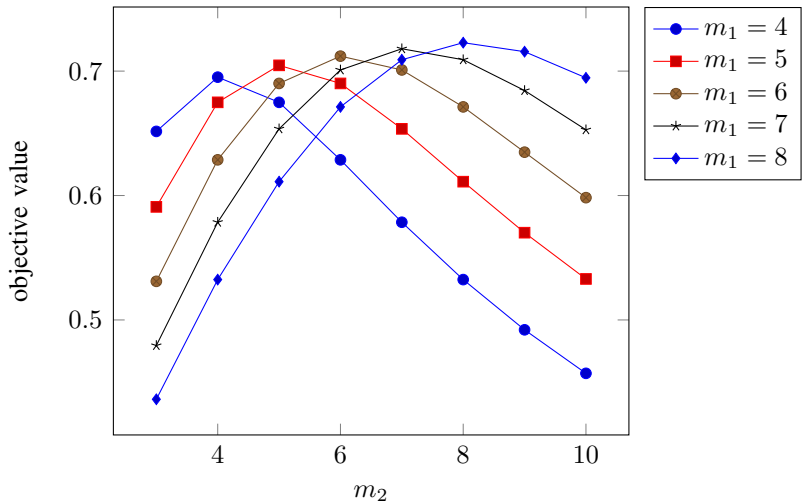
\begin{figure}
  \centering
  \begin{tikzpicture}
    \begin{axis}[font=\small,
      xlabel=$m_2$, ylabel=objective value,
      legend pos=outer north east
      ]
      \addplot coordinates {
        (3, 0.6515432826700548)(4, 0.6951462712499902)(5, 0.6749022217064367)(6, 0.6286888320043499)(7, 0.5785282678875476)(8, 0.5324256066141568)(9, 0.4920671210563124)(10, 0.45708108200970227)
      };
      \addlegendentry{$m_1=4$}
      \addplot coordinates {
        (3, 0.5908776872530342)(4, 0.6749022217064367)(5, 0.7046350070467768)(6, 0.6901017880348363)(7, 0.6535932594543027)(8, 0.6111332189426452)(9, 0.5701275179668231)(10, 0.5329538416103055)
      };
      \addlegendentry{$m_1=5$}
      \addplot coordinates {
        (3, 0.5310285837100782)(4, 0.6286888320043499)(5, 0.6901017880348363)(6, 0.712021492331114)(7, 0.7009344003786235)(8, 0.6712319484438797)(9, 0.634843040912084)(10, 0.5982788624052231)
      };
      \addlegendentry{$m_1=6$}
      \addplot coordinates {
        (3, 0.4794394685764624)(4, 0.5785282678875476)(5, 0.6535932594543027)(6, 0.7009344003786235)(7, 0.7179581161898583)(8, 0.70913470026034)(9, 0.6843925994025323)(10, 0.6528315321171955)
      };
      \addlegendentry{$m_1=7$}
      \addplot coordinates {
        (3, 0.4362310363635315)(4, 0.5324256066141568)(5, 0.6111332189426452)(6, 0.6712319484438797)(7, 0.70913470026034)(8, 0.7228535235969599)(9, 0.7156117585847128)(10, 0.6946042301472493)
      };
      \addlegendentry{$m_1=8$}
  \end{axis}    
  \end{tikzpicture}
  \caption{Illustration of the objective of \eqref{eq:mmlog}. For each $m_1=4,\ldots,8$, we plot the objective values for $m_2=3,\ldots,10$.}
  \label{fig:ox}
\end{figure}

Note that both our approach and the algorithm in \cite{zhou19} have an exponential complexity in terms of the flow length. A seemingly intuitive way to solve \eqref{eq:mm} with a low computational cost is to alternately optimize the recoding number of every single link while fixing the others until no link can be updated to bring improvement. Here, we give an example to illustrate that this intuition may have a poor performance.
Consider a two-hop flow where both links have capacity $ 1 $ and loss rate $ 0.2 $. Assume the packet losses are independent and we use natural logarithm as the utility.
The objective of \eqref{eq:mmlog} is a function of the recoding numbers $m_1$ and $m_2$ (where the superscript is dropped for conciseness). We plot the objective function in Fig.~\ref{fig:ox}. Suppose we start with $m_1=5$ and optimize $m_2$. From the figure, we can see that the best value of $m_2$ is $5$. Now we fix $m_2=5$ and optimize $m_1$. Then, we get $ m_1=5 $, which implies that the algorithm stops when $(m_1,m_2)=(5,5)$. 
However, our local search algorithm can find $(m_1,m_2)= (6,6) $ in $ \mc N((5,5)) $ which is a better solution.

\subsection{Numerical Results}
\label{sec:nr1}

To evaluate the performance of the dual-based algorithm for \eqref{eq:ur}, we solve the problem for several instances. We use a network with node set $\mc V= \{v_0,\ldots, v_8\}$ and edge set $\mc E = \{e_i=(v_{i-1},v_i), i=1,\ldots,8\}$. 
Let $\epsilon_i$ be the packet loss rate of $e_i$ and $c_i$ be the communication rate of $e_i$. We test $11$ cases of different settings as listed in Table~\ref{tab:1}, where without otherwise specified, $\epsilon_i=0.2$ and $c_i = 1$.
We use the two-hop interference model~\cite{ephremides1990scheduling}. Here we assume the packet losses are independent and the packet loss model $q_e$ can be calculated by \eqref{eq:il}.
For BATS codes, we use batch size $M=16$ and base field size $q=256$ for all the flows.

Although we can use general utility functions in our algorithm, we use the natural logarithm ($\log$) in our numerical evaluations.
Logarithm utility functions imply the proportional fairness criterion~\cite{kelly1997charging} which have been used in the analysis of TCP~\cite{low2017analytical}. See more about the choices of utility functions in~\cite{srikant2004mathematics, low2017analytical}.

\begin{table}[t]
    \centering
    \caption{Network and Flow Settings. In the ``flows'' column, $[11111000;00111111]$  means that the first flow contains the links  $ e_1 $ to $ e_5$ and the second flow contains the links $ e_3 $ to $ e_8 $. In the ``loss rate'' and ``commun. rate'' columns, without otherwise specified, $\epsilon_i=0.2$ and $c_i=1$. Here $ \epsilon_{3,4,5}=0.1 $ means $ \epsilon_3=\epsilon_4=\epsilon_5 =0.1 $, and $ c_{3,4,5}=2 $ means $ c_3=c_4=c_5=2 $.}\label{tab:1}
    \begin{tabular}{|l|l|l|l|}
      \hline
      case & \multicolumn{1}{|c|}{flows} & \multicolumn{1}{|c|}{loss rate} & \multicolumn{1}{|c|}{commun. rate} \\ \hline
      \multicolumn{1}{|c|}{1} &    \multicolumn{1}{|c|}{$[11111000;00111111]$}      &     \multicolumn{1}{|c|}{/}      &     \multicolumn{1}{|c|}{/}       \\ \hline
      \multicolumn{1}{|c|}{2}                       &  \multicolumn{1}{|c|}{same as case 1}     &      \multicolumn{1}{|c|}{/}       &    \multicolumn{1}{|c|}{$ c_{3,4,5}=2 $ }     \\ \hline
      \multicolumn{1}{|c|}{3}                      &    \multicolumn{1}{|c|}{same as case 1}  &      \multicolumn{1}{|c|}{ / }         &       \multicolumn{1}{|c|}{ $ c_{1,2,6,7,8}=1/2 $ }       \\ \hline
      \multicolumn{1}{|c|}{4}                      &    \multicolumn{1}{|c|}{same as case 1}  &      \multicolumn{1}{|c|}{ / }         &       \multicolumn{1}{|c|}{ $ c_{1,2,6,7,8}=1/4 $ }       \\ \hline
      \multicolumn{1}{|c|}{5 }                      &      \multicolumn{1}{|c|}{same as case 1}       &   \multicolumn{1}{|c|}{ $ \epsilon_{3,4,5}=0.1 $  }     &      \multicolumn{1}{|c|}{/}      \\ \hline
      \multicolumn{1}{|c|}{6}                      &   \multicolumn{1}{|c|}{same as case 1}          &     \multicolumn{1}{|c|}{$ \epsilon_{3,7}=0.1 $  }    &     \multicolumn{1}{|c|}{/}        \\ \hline
      \multicolumn{1}{|c|}{7     }                  &    \multicolumn{1}{|c|}{same as case 1 }  &      \multicolumn{1}{|c|}{ $ \epsilon_{1,2,6,7,8}=0.1 $ }         &       \multicolumn{1}{|c|}{ / }       \\ \hline
      \multicolumn{1}{|c|}{8  }                     &    \multicolumn{1}{|c|}{same as case 1 }  &      \multicolumn{1}{|c|}{ $ \epsilon_{1,2,6,7,8}=0.4 $ }         &       \multicolumn{1}{|c|}{ / }       \\ \hline
      \multicolumn{1}{|c|}{9    }                   &   \multicolumn{1}{|c|}{$[11111111;11111111]$ }     &        \multicolumn{1}{|c|}{ / }       &       \multicolumn{1}{|c|}{ / }       \\ \hline
      \multicolumn{1}{|c|}{10  }                    &    \multicolumn{1}{|c|} { $ [11111111;00111111]$ }  &      \multicolumn{1}{|c|}{ / }         &     \multicolumn{1}{|c|}{ / }         \\ \hline
      \multicolumn{1}{|c|}{11  }                    &    \multicolumn{1}{|c|}{ $ [11111111;00111100]$ }  &      \multicolumn{1}{|c|}{ / }         &       \multicolumn{1}{|c|}{ / }       \\ \hline
    \end{tabular}
  \end{table}

The numerical results of solving these $11$ cases are given in Table~\ref{tab:2} and~\ref{tab:3}.
Table~\ref{tab:2} includes the recoding numbers and batch rates obtained using the dual-based algorithm for solving \eqref{eq:ur}.
In Table~\ref{tab:3},  $U_1=\log(\alpha_1\E[\bh^1])$ and $U_2=\log(\alpha_2\E[\bh^2])$ are the utilities of the two flows obtained using the dual-based algorithm for solving \eqref{eq:ur}, and $\tilde U$ is the optimal value of \eqref{eq:3}, which provides an upper bound of $ U \triangleq U_1+U_2$. We see that the utilities of the two flows  are very close to each other, which means that the flow control induced by the algorithm is fair.  

 Denote by $\tilde U_1$ and $\tilde U_2$ the optimal utilities of the two flows of \eqref{eq:3} respectively. Then, $\tilde U = \tilde U_1+\tilde U_2$ is the optimal value of \eqref{eq:3}.
To compare the achievable utilities of \eqref{eq:ur} and \eqref{eq:3}, we define the \emph{utility ratio} 
\begin{equation}\label{eq:kappa}
\kappa = \sqrt{\frac{e^{U_1}}{e^{\tilde U_1}}\frac{e^{U_2}}{e^{\tilde U_2}}} = e^{(U-\tilde U)/2}.
\end{equation}
In other words, $\kappa$ is the geometric average of the throughput ratios of all flows.\footnote{For a network of $k$ flows, we can extend the definition as $\kappa = e^{(U-\tilde U)/k}$.}
As another interpretation of $\kappa$, we scale $\alpha_i$ by $1/\kappa$ and then evaluate the total utility:
\begin{IEEEeqnarray*}{rCl}
  \log( \alpha_1\overline{\bh^1}/\kappa) + \log( \alpha_2\overline{\bh^2}/\kappa) & = & -2 \log(\kappa) + U = \tilde U.
\end{IEEEeqnarray*}
From this point of view, $\kappa$ is the rate scaling factor so that the solution can achieve the cut-set bound.
We know that $\kappa \le 1$ as $U\leq \tilde U$.
In the last column of Table~\ref{tab:3}, we show $\kappa$ for all the cases and observe that  $ \kappa $ is at least $ 83.86\% $ for our evaluated cases.

\begin{table}[]
  \centering
  \caption{Recoding Numbers for Random Linear Recoding Obtained by the Dual-Based Algorithm}\label{tab:2}
  \begin{tabular}{|l|l|l|l|l|l|l|l|}
    \hline
    case & \multicolumn{1}{|c|}{$\alpha_1\times 10^2 $}& \multicolumn{1}{|c|}{$ \alpha_2 \times 10^2$}& \multicolumn{1}{|c|}{$ {m}^1 $} &\multicolumn{1}{|c|}{$ {m}^2 $} \\ \hline
    \multicolumn{1}{|c|}{1} & $ 0.877 $   &   $ 0.877 $  & \multicolumn{1}{|c|}{$[32, 31, 19, 19, 19] $}&   \multicolumn{1}{|c|}{$[19, 19, 19, 29, 33, 31] $}  \\ \hline
    \multicolumn{1}{|c|}{2} &  $ 1.645 $   &   $ 1.606 $  & \multicolumn{1}{|c|}{$ [21,20,20,21,21] $}& \multicolumn{1}{|c|}{ $[20,20,21,20,21,21]$ } \\ \hline
    \multicolumn{1}{|c|}{3}&  $ 0.819 $   &  $ 0.794 $    & \multicolumn{1}{|c|}{$[20,20,20,21,21]$} & \multicolumn{1}{|c|}{$[20, 21, 21, 20, 21, 22]$} \\ \hline
    \multicolumn{1}{|c|}{4}&  $ 0.525 $  &   $ 0.417 $    & \multicolumn{1}{|c|}{$[19, 19, 21, 25, 24]$} & \multicolumn{1}{|c|}{$[22, 26, 24, 20, 20, 20]$}  \\ \hline
    \multicolumn{1}{|c|}{5}&  $ 0.980 $   &    $ 0.980 $   & \multicolumn{1}{|c|}{$[30,28,17,17,17]$}& \multicolumn{1}{|c|}{$[17,17,17,32,30,28]$} \\ \hline
    \multicolumn{1}{|c|}{6}&   $ 0.909 $   &   $ 0.909 $    &\multicolumn{1}{|c|}{$[32,28, 17, 19, 19] $}  &\multicolumn{1}{|c|}{ $[17, 19, 19, 30, 27,28]$}  \\ \hline
    \multicolumn{1}{|c|}{7}&   $ 0.877 $   &    $ 0.877 $  &\multicolumn{1}{|c|}{$[28,25, 19, 19, 19]$} & \multicolumn{1}{|c|}{$[19, 19, 19, 26,32,24]$}   \\ \hline
    \multicolumn{1}{|c|}{8}&  $ 0.877 $   &  $ 0.877 $    &\multicolumn{1}{|c|}{$[36,35,19,19,19]$} & \multicolumn{1}{|c|}{$[19,19,19,36,36,36]$}   \\ \hline
    \multicolumn{1}{|c|}{9}&   $ 0.769 $     &  $ 0.769 $    &\multicolumn{1}{|c|}{$[22, 22, 21, 22, 22, 21, 22, 22]$} &\multicolumn{1}{|c|}{$[22, 22, 21, 22, 22, 21, 22, 22]$}   \\ \hline
    \multicolumn{1}{|c|}{10}&   $ 0.794 $    &   $ 0.794 $   &\multicolumn{1}{|c|}{$[30,26,21,21,21,21,21,21]$} &\multicolumn{1}{|c|}{$[21, 21, 21, 21, 21, 21]$}  \\ \hline
    \multicolumn{1}{|c|}{11}& $ 0.862 $   &  $ 0.862 $   &\multicolumn{1}{|c|}{$[27,31,20,19,19,20,27,31]$} & \multicolumn{1}{|c|}{$[20,19,19,20]$}  \\ \hline
  \end{tabular}
\end{table}

\begin{table}[]
  \centering
  \caption{Comparison Table for Random Linear  Recoding}\label{tab:3}
  \begin{tabular}{|c|c|c|c|c|}
    \hline
     case & 	\multicolumn{1}{|c|}{$U_1$} &  $U_2 $ & 	\multicolumn{1}{|c|}{$\tilde U$} & $\kappa$ \\ \hline
    \multicolumn{1}{|c|}{1}  &  \multicolumn{1}{|c|}{$ -2.119$ }      &   \multicolumn{1}{|c|}{$ -2.119$}      &     \multicolumn{1}{|c|}{$ -4.030 $}        &        \multicolumn{1}{|c|}{$ 90.12\% $}     \\ \hline
    \multicolumn{1}{|c|}{2 }  &\multicolumn{1}{|c|}{$ -1.452$}       &   \multicolumn{1}{|c|}{$ -1.495 $}      &   \multicolumn{1}{|c|}{$ -2.644     $  }    &      \multicolumn{1}{|c|}{$ 85.94\% $}       \\ \hline
    \multicolumn{1}{|c|}{3 } &\multicolumn{1}{|c|}{$ -2.159 $}        &   \multicolumn{1}{|c|}{  $ -2.186 $}    &    \multicolumn{1}{|c|}{$ -4.030 $}         &      \multicolumn{1}{|c|}{$ 85.43\% $}       \\ \hline
    \multicolumn{1}{|c|}{4}  &\multicolumn{1}{|c|}{$-2.610$}       &    \multicolumn{1}{|c|}{$ -2.821 $}     &     \multicolumn{1}{|c|}{$ -5.215 $   }     &   \multicolumn{1}{|c|}{$ 89.76\% $}          \\ \hline
    \multicolumn{1}{|c|}{5}  &\multicolumn{1}{|c|}{$ -1.969$}        &    \multicolumn{1}{|c|}{$ -1.969 $}     &    \multicolumn{1}{|c|}{$ -3.794 $}         &     \multicolumn{1}{|c|}{$ 93.05   \% $}        \\ \hline
    \multicolumn{1}{|c|}{6}  &\multicolumn{1}{|c|}{$ -2.071$}       & \multicolumn{1}{|c|}{$ -2.071 $} &    \multicolumn{1}{|c|}{$-3.954  $ }        &     \multicolumn{1}{|c|}{  $ 91.03\% $}      \\ \hline
    \multicolumn{1}{|c|}{7}  &\multicolumn{1}{|c|}{$-2.119$}      &   \multicolumn{1}{|c|}{$ -2.119 $}      &    \multicolumn{1}{|c|}{$ -4.030 $}         &     \multicolumn{1}{|c|}{ $ 90.12\% $}       \\ \hline
    \multicolumn{1}{|c|}{8}  &\multicolumn{1}{|c|}{$ -2.120 $}      &   \multicolumn{1}{|c|}{$ -2.120 $}      &       \multicolumn{1}{|c|}{$ -4.030 $}      &     \multicolumn{1}{|c|}{$ 90.03\% $}        \\ \hline
    \multicolumn{1}{|c|}{9} &\multicolumn{1}{|c|}{$ -2.191$}       &   \multicolumn{1}{|c|}{$ -2.191 $}      &      \multicolumn{1}{|c|}{$ -4.030 $}       &   \multicolumn{1}{|c|}{$ 83.86\% $}          \\ \hline
    \multicolumn{1}{|c|}{10} &\multicolumn{1}{|c|}{$ -2.172$}       &   \multicolumn{1}{|c|}{$ -2.172 $}      &    \multicolumn{1}{|c|}{$ -4.030 $}         &   \multicolumn{1}{|c|}{$ 85.47\% $}          \\ \hline
    \multicolumn{1}{|c|}{11} &\multicolumn{1}{|c|}{$ -2.137$}       &  \multicolumn{1}{|c|}{$ -2.137 $}       &     \multicolumn{1}{|c|}{$ -4.030 $ }       &    \multicolumn{1}{|c|}{$ 88.51\% $}         \\ \hline
  \end{tabular}
\end{table}

\section{Algorithms for General Adaptive Recoding Problem}
\label{sec:algorithm_ad}

In this section, we discuss the algorithms for solving the general adaptive recoding problem~\eqref{eq:1}. Following the discussion in the last section, one may consider extending the dual-based algorithm by replacing  $m_e^i$ with $p_e^i$ so that~\eqref{eq:update_recoding} becomes the optimizer of
\begin{equation}\label{eq:update_recoding_g}
  \max_{\alpha_i, (p_e^i,e\in \mc P^i) } \left[U_i(\alpha_i\E[h^i])-\alpha_i \sum_{e\in \mc P^i} q_e(t) \overline m_e^i \right].
  \end{equation}
As $p_e^i$ are continuous variables, the approach in Section~\ref{sec:rnu} cannot be directly extended to update $p_e^i$ in~\eqref{eq:update_recoding_g}.
We may try to solve \eqref{eq:update_recoding_g} by gradient search, which is similar to the primal-dual algorithm for classical NUM problems without network coding~\cite{shakkottai2008network}. However, due to the nonconcavity of the problem, our experiments show that it is hard to escape a local optimum with a poor objective value. We leave the formulations of the primal-dual algorithm in Appendix~\ref{sec:primal_dual}. %
In the following, we discuss a two-step approach for solving \eqref{eq:1} with adaptive recoding. %

\subsection{Two-step Approach}
We solve \eqref{eq:1} in two steps. In the first step, we solve the nonadaptive recoding version  \eqref{eq:ur} of \eqref{eq:1} using the dual-based algorithm discussed in Section~\ref{sec:algorithm_u}. Denote by $((\alpha_i,(m_e^i,e\in \mc P^i))_i,s)$ the optimizer of \eqref{eq:ur}.
In the second step, we reallocate the recoding resources to gain the advantage of adaptive recoding. In the nonadaptive recoding optimization, all batches from flow $\mc P^i$ have the same $m_e^i$ recoded packets transmitted on edge $e$. With adaptive recoding, the number of recoded packets for the batches of relatively lower ranks can be reduced so that some bandwidth along the path of the flow is released. We consider two strategies to reallocate the released bandwidth:
\begin{enumerate}[i)]
	\item Transmit more recoded packets for the batches of relatively higher ranks; or
	\item Transmit more batches, i.e., transmit in a higher batch rate.
\end{enumerate}

To manage the computational cost, we reallocate the recoding resources of each flow $\mc P^i$ separately. Suppose $\mc P^i=\{e_\ell^i=(v_{\ell-1}^i,v_{\ell}^i),\ell=1,\ldots,L^i\}$. For each $\ell = 1,\ldots,L^i$, we write $ p_{e_{\ell}^i}^i $ as $ p_{\ell}^i $ to simplify the notation. Let $\eta_i \geq 1$ be the variable for scaling the batch rate $\alpha_i$. 
With the scaled batch rate $ \eta_i \alpha_i $, we can find  a  feasible solution $ (p_e^i, i\in \mc P^i) $ of \eqref{eq:1} by solving 
\begin{equation}\label{eq:xkk}
  \begin{IEEEeqnarraybox*}[][c]{rCl}
  \max_{p_{\ell}^i} & \quad & \E\left[\bh_{\ell}^i\right] \\
  \text{s.t.} & & \overline{m}_{e_{\ell}^i}^i \leq m_{e_{\ell}^i}^i/\eta_i,
\end{IEEEeqnarraybox*}
\end{equation}
where $\bh^i_{\ell}$ is the rank distribution of the batches received at node $v_{\ell}^i$ in flow $\mc P^i$, and $\overline{m}_{e_\ell^i}^i= \sum_{m,r} m p_{\ell}^i(m|r) \bh_{\ell-1}^i(r)$.   In the objective function, $\bh_{\ell}^i$ can be obtained from $\bh_{\ell-1}^i$ using the transition matrix studied in Section~\ref{sec:ana}.

Suppose $\bh_{0}^i=(0,\ldots,0,1)$, i.e., all the batches at the source node have the maximum rank. For a given $\eta_i$, the optimization problem \eqref{eq:xkk} with $\ell=1$ can be solved by using the algorithm proposed in~\cite{yin_adaptive19}. Hence, we obtain $\bh_{1}^i$. Similarly, we can solve \eqref{eq:xkk} for $\ell=2,\ldots,L^i$ sequentially.
Let $R_i(\eta_i)$ be the optimal value of \eqref{eq:xkk} for $\ell=L^i$.
We can then tune $\eta_i$ to maximize $\eta_i \alpha_i R_i(\eta_i)$ for $\eta_i\geq 1$.

Note that the solution of $p_{\ell}^i$ obtained in the above process is almost deterministic if the packet loss pattern is a stationary stochastic process~\cite{yin_adaptive19}. 
For online implementation, the optimization problem \eqref{eq:xkk} can be solved at node $v_{\ell-1}^i$, which is the network node that needs $p_{\ell}^i$ for recoding. The distribution $\bh_{\ell-1}^i$ can be calculated at node $v_{\ell-1}^i$ by counting the numbers of batches of different ranks in flow $\mc P^i$. Therefore, \eqref{eq:xkk} can be solved locally in an upstream-downstream order.

\begin{table*} %
  \centering
    \caption{Performance evaluations of adaptive recoding using the two-step algorithm with independent packet loss.}\label{tab:4}
    \begin{tabular}{|c|c|c|c|}
      \hline 
      & \multicolumn{3}{c|}{two-step algorithm } \\ \hline
      \rule{0pt}{8pt}
      case & $ U_1$&$U_2 $        &      $\kappa$     \\ \hline
      
      \multicolumn{1}{|c|}{1}    &     $-2.095$&$-2.095$       &    $92.33\%$              \\ \hline
      
      \multicolumn{1}{|c|}{2}    &           $-1.426$&$-1.467$      &       $ 88.30\%$          \\ \hline
      
      \multicolumn{1}{|c|}{3}    &        $-2.119$&$-2.160$     &       $ 88.28\%$          \\ \hline
      
      \multicolumn{1}{|c|}{4}   &      $-2.587$&$-2.794$   &      $ 92.04\%$              \\ \hline
      
      \multicolumn{1}{|c|}{5}   &         $-1.951$&$-1.951$      &      $ 94.71\%$               \\ \hline
      
      \multicolumn{1}{|c|}{6}    &         $-2.054$&$-2.054$     &     $92.57 \%$                  \\ \hline
      
      \multicolumn{1}{|c|}{7}   &       $-2.095$&$-2.095$      &     $92.33 \%$                \\ \hline
      
      \multicolumn{1}{|c|}{8}   &         $-2.095$&$-2.095$     &       $ 92.28\%$             \\ \hline
      
      \multicolumn{1}{|c|}{9}   &        $-2.165$&$-2.165$      &      $86.06 \%$               \\ \hline
      
      \multicolumn{1}{|c|}{10}   &         $-2.146$&$-2.145$          &      $87.77 \%$         \\ \hline
      \multicolumn{1}{|c|}{11}    &           $-2.111$&$-2.111$     &     $ 90.83\%$        \\ \hline
  \end{tabular}
\end{table*}

\subsection{Numerical Results}
\label{sec:nr}

In this section, we evaluate the optimization algorithms for solving the adaptive recoding problem~\eqref{eq:1} with natural logarithms as the utilities. We consider the same network and flow settings and also the same BATS code parameters as in Section~\ref{sec:nr1}.

\subsubsection{Independent Packet Loss Model}
We first consider the case that packet losses are independent. 
For each case listed in Table~\ref{tab:1}, we  solve \eqref{eq:1} by using the two-step algorithm.
The utilities of the two-step algorithm are given in Table~\ref{tab:4}.
We also give the values of $\kappa$ for comparison.

From Table~\ref{tab:4}, we can see that $U_1$ and $U_2$ are all very close to each other in each case, which implies the fairness of the flow control.
In each case, we observe that $U_1$, $U_2$ and $\kappa$ given by the two-step algorithm are better than those given by the nonadaptive recoding algorithm shown in Table~\ref{tab:3}.
In terms of $ \kappa $, the two-step algorithm is about $1.7\%$ to $2.75 \%$ better than the nonadaptive recoding algorithm.
The relatively small gain of using adaptive recoding for independent packet loss is expected as the ranks of the batches are highly concentrated.

\begin{table*}[htbp]
  \centering
  \caption{Performance evaluation for Gilbert-Elliott packet loss model}\label{tab:5}
  \begin{tabular}{|c|c|c|c|c|c|c|}
    \hline
    & \multicolumn{3}{c|}{nonadaptive recoding } & \multicolumn{3}{c|}{adaptive recoding: two-step approach } \\ \hline
    \rule{0pt}{8pt}
    case & $U_1$&$U_2$    &     $\kappa$    &  $U_1$&$U_2$    &      $\kappa$       \\ \hline
    
    \multicolumn{1}{|c|}{1}    &        $-2.290$&$-2.289$    & \multicolumn{1}{|c|}{ $76.01\% $ }     &     $-2.233 $&$-2.231  $       &    $  80.50 \%$             \\ \hline
    
    \multicolumn{1}{|c|}{2}    &       $-1.718$&$ -1.754 $    &  \multicolumn{1}{|c|}{$  66.10\% $  }  &        $-1. 627  $&$-1. 681  $    &       $  71.74 \%$        \\ \hline
    
    \multicolumn{1}{|c|}{3}    &         $ -2.408$&$ -2.448 $   &   \multicolumn{1}{|c|}{$ 66.16\%  $}   &        $-2. 313 $&$-2.373   $  &              $  72.03  \%$        \\ \hline
    
    \multicolumn{1}{|c|}{4}    &         $ -2.795$&$ -3.059$    &  \multicolumn{1}{|c|}{ $  72.66\% $}   &      $-2.715$&$-2.967$    &      $ 79.17   \%$       \\ \hline
    
    \multicolumn{1}{|c|}{5}   &  $ -2.046 $ &$ -2.044 $ & $ 86.23\% $  &  $ -2.001 $ & $ -2. 003$ & $  90.02 \% $  \\ \hline
    
    \multicolumn{1}{|c|}{6}   &  $ -2.221 $ &$ -2.223 $ & $ 78.30\% $  &  $ -2.190 $ & $ -2.190 $ & $80.84 \% $  \\ \hline
    
    \multicolumn{1}{|c|}{7}   &  $ -2.282 $ &$ -2.283 $ & $ 76.55\% $  &  $ -2. 231$ & $ -2. 231$ & $ 80.60\% $  \\ \hline		
    
    \multicolumn{1}{|c|}{8}   &  $ -2.356 $ &$ -2.369 $ & $ 70.63\% $  &  $ -2.252 $ & $ -2.274 $ & $78.07 \% $  \\ \hline		
    
    \multicolumn{1}{|c|}{9}   &  $ -2.470$ &$ -2.470 $ & $ 63.43\% $  &  $ -2.411 $ & $ -2. 411$ & $ 67.31\% $  \\ \hline
    
    \multicolumn{1}{|c|}{10}   &  $ -2.436 $ &$ -2.435 $ & $ 65.65\% $  &  $ -2.364 $ & $ -2.364 $ & $70.52 \% $  \\ \hline
    
    \multicolumn{1}{|c|}{11}   &  $ -2.356 $ &$ -2.346 $ & $ 71.48\% $  &  $ -2.264 $ & $ -2. 265$ & $77.92 \% $  \\ \hline
  \end{tabular}
\end{table*}

\subsubsection{Gilbert-Elliott Packet Loss Model}
\label{subsec:GEnum}
      
The Gilbert-Elliott (GE) model \cite{gilbert1960capacity,elliott1963estimates} has been
widely applied to describe burst error patterns in communication channels. Here, we describe a GE model for packet loss on a network link. A GE model has two states $ G $ (good) and $ B $ (bad).  In state $ G $ (resp. $B$), a packet transmitted on this link can be correctly received with probability $ s_G $ (resp. $s_B$). The state transition probability from state $ G $ to state $ B $ is $ p_{G\to B} $ and the one from state $ B $ to state $ G $ is $ p_{B\to G}$. We use $  \mathrm{GE}(s_G,s_B,p_{G\to B},p_{B\to G})$ to denote the GE model with parameters $ s_G,s_B,p_{G\to B},p_{B\to G} $.
Let $(\pi_G, \pi_B)$ be the steady state of the GE model. We have
\begin{equation*}
	\pi_G=\frac{p_{B\to G}}{p_{G\to B}+p_{B\to G}}, \quad \pi_B=\frac{p_{G\to B}}{p_{G\to B}+p_{B\to G}}.
\end{equation*}
The (average) loss rate of the GE model is $1 - \pi_Gs_G - \pi_Bs_B$. 

For each link in the network settings shown in
Table~\ref{tab:1}, we use a GE model with the corresponding loss rate. $\mathrm{GE}(1,0.8,10^{-3},10^{-3}) $, $ \mathrm{GE}(1,0.6,10^{-3},10^{-3}) $ and $ \mathrm{GE}(0.8,0.4,10^{-3},10^{-3})$ are used for loss rates $ 0.1 $, $ 0.2 $ and $ 0.4 $, respectively. 
Our NUM problems need the batch-wise packet loss model $q_e(r|m)$ for each link $e$. For each $m=1,\ldots, 100$, we use the empirical distribution of the number of received packets when transmitting $m$ packets as $q_e(\cdot|m)$. The number of samples is $ 10000 $.

With this preparation, the NUM problem \eqref{eq:1} associated with each network setting with the GE model is ready to be solved by the two-step algorithm. In Table~\ref{tab:5}, we give the utilities of both flows in the first and the second steps, and we observe that our algorithm can achieve the fairness as well. Note that the corresponding problem \eqref{eq:3} of \eqref{eq:1} is the same as the one with the independent loss model, and hence the cut-set upper bound on the total utility is the same.
For each case, the values of $\kappa$ for both steps are given in Table~\ref{tab:5}. 
These values are
about $ 6\%$ to $10.5\% $ higher than the corresponding nonadaptive solution in the first step for eight of the cases.
Comparing with the results about independent packet loss, we observe that adaptive recoding with GE packet loss models has a larger gain than that with independent packet loss models.

\section{Simulation Results}
\label{sec:sr}

In the last section, we only solve the  optimization problems numerically. %
To observe how the results of these algorithms work in real network communications, we use ns-3~\cite{Riley2010} to implement a simulator with BATS codes, which supports both the independent packet loss model and the GE packet loss model. For a rate vector $s\in \cov(\mc R)$, we can obtain the corresponding scheduling so that the simulator can operate the network without collision~\cite{lin2006tutorial}. 

Our process of using the simulator is as follows.
\begin{itemize}
\item We first configure the simulator using the setting of case $ 1 $ in Table~\ref{tab:1}, and set up either the independent packet loss model or the GE packet loss model.
The batch-wise packet loss model $q_e(\cdot|m)$ is obtained as the empirical distribution of the number of received packets when transmitting $m$ packets, for each $ m=1,\ldots,100 $. The number of samples for each $m$ is $ 10000 $.
\item We then solve the corresponding NUM problem~\eqref{eq:1} using the two-step algorithm, which returns the coding parameters $(\alpha_i,(m_e^i,e\in \mc P^i))$ for each flow $\mc P^i$ and the scheduling rate vector $s\in \cov(\mc R)$.
\item  At last, we substitute these parameters into the simulator and perform the simulation.
\end{itemize}

During the simulation, we keep track on the buffer size, defined as the number of recoded packets of all the flows which are generated but not yet transmitted, at each network node $v_0, v_1,\ldots, v_7$.
We also record the empirical rank distribution at the destination node of each flow.

\begin{figure}
	\centering
	\includegraphics[width=\textwidth]{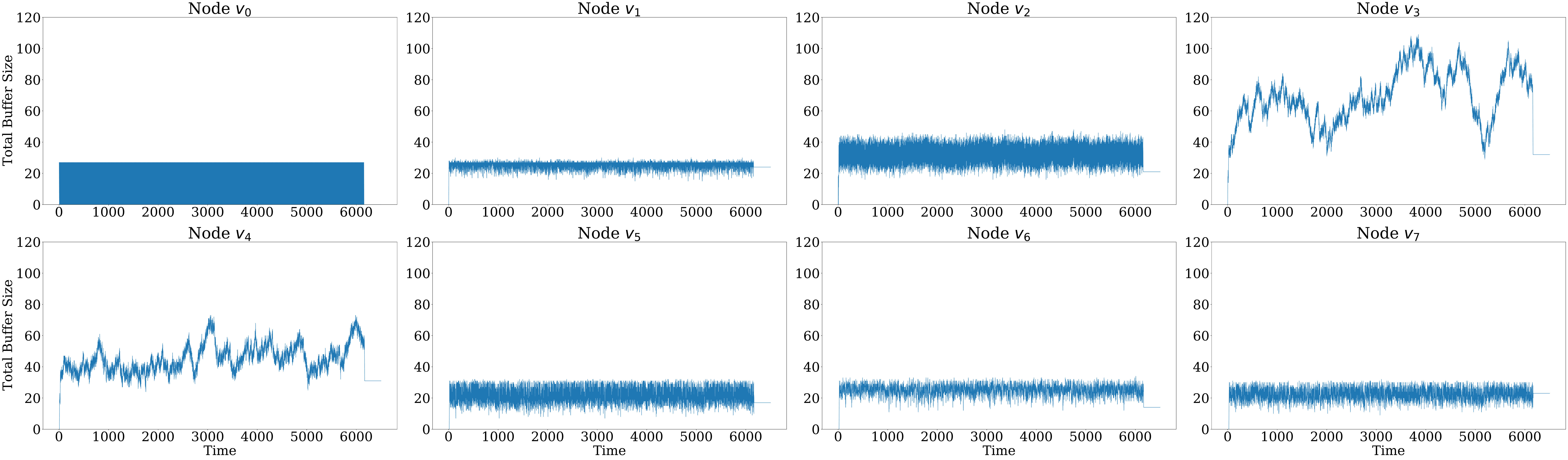}
	\caption{The buffer size at each node for the independent packet loss model.}
	\label{fig:1}
\end{figure}

When all links have independent packet loss, the utilities of the two flows obtained in the simulation are $-2.094 $ and $ -2.104 $ respectively, which are very close to the utilities obtained by solving~\eqref{eq:1} using the two-step algorithm (see Table~\ref{tab:4}). The buffer sizes are counted at each timeslot of the simulation. Fig.~\ref{fig:1} shows the instantaneous buffer sizes at $v_0, v_1,\ldots,v_7 $ at each timeslot. %
We can see that the buffer sizes do not grow indefinitely at all the nodes.
At the source node of each flow, batches are generated according to the given batch rate. Hence the buffer size jumps between $0$ and the recoding number periodically (as illustrated by the figure for $v_0$).
The dynamics of the buffer sizes at the other nodes can be classified into two classes. The first class includes nodes $v_1,v_2,v_5,v_6,v_7$, where the buffer sizes are within $20$ and $40$ most of the time. The second class includes nodes $v_3$ and $v_4$, where the buffer sizes may change dramatically over time.
If we substitute the solution obtained by the two-step algorithm back into \eqref{eq:1} and check the inequality constraints, we can observe that the inequality constraints for the outgoing links of $ v_2,v_3 , v_4 $ are saturated, while those for the going links of $ v_1,v_5,v_6,v_7 $ are not.
Node $v_2$ is the source node of $\mc P^2$ and an intermediate node of $\mc P^1$. At node $v_2$, the recoded packets periodically generated for $\mc P^2$ increase its buffer size, thus giving the curve for $v_2$ a relatively more regular shape than those of $v_3$ and $v_4$.

\begin{figure}
	\centering
	\includegraphics[width=\textwidth]{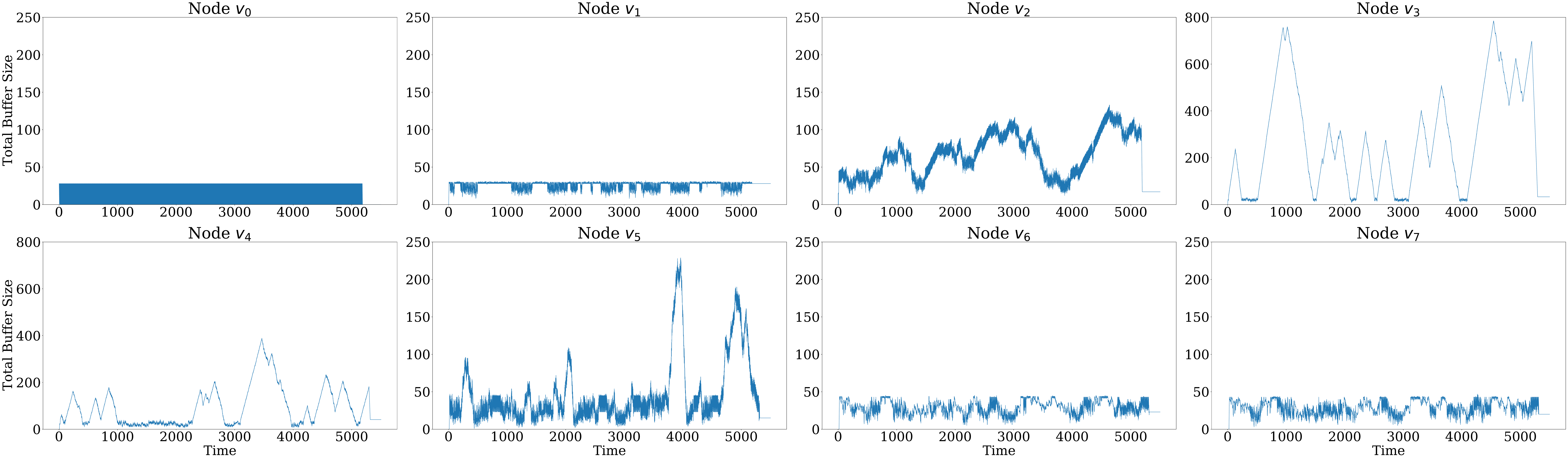}
	\caption{The buffer size at each node for the GE packet loss loss model.}
	\label{fig:2}
\end{figure}

When all links have GE packet loss, we use the same model setting as in Section~\ref{subsec:GEnum} for case $ 1 $. The utilities of the two flows obtained in the simulation are $ -2.307 $ and $ -2.288 $ respectively, which are again very close to the values obtained by solving~\eqref{eq:1} using the two-step algorithm (see Table~\ref{tab:5}).
From Fig.~\ref{fig:2}, we observe that the buffer sizes at $ v_0,\ldots,v_7 $ during the simulation do not grow indefinitely. Except at $v_0$, the buffer sizes at all the other nodes change more dramatically than the independent loss simulation.  
Nodes $ v_1,v_6,v_7 $ have buffer sizes smaller than $ 50 $, and nodes $ v_2,v_5 $ have buffer sizes smaller than $100$  most of the time.
The buffer sizes at nodes $ v_3,v_4 $ can be close to $800$ and $400$, respectively. 
If we substitute the solution obtained by the two-step algorithm back into \eqref{eq:1} and check the inequality constraints, we can observe that the inequality constraints for the outgoing links of $ v_2,v_3 , v_4 $ are saturated, the one for the outgoing link of $ v_5 $ is almost saturated, and those for the outgoing links of $ v_1,v_6,v_7 $ are not saturated.

\section{Concluding Remarks}
\label{sec:remark}

We introduced a network utility maximization (NUM) problem for a multi-flow network employing BATS codes with adaptive recoding, where a batch-wise packet loss model is used to capture the packet loss events on the network links.
We proposed some preliminary algorithms for solving the problem and illustrated how to evaluate and compare the performance of the algorithms. Although we only discussed the NUM problem for BATS codes, our discussion can be applied to other batched network codes where adaptive recoding can be adopted and the performance of the outer code can be measured by the expected rank of the batch transfer matrices. 

The algorithms provided in this paper are for solving the NUM problem with all the required network statistics known in a centralized manner. In the scenarios where the network topology, flow settings and link statistics are stable for a long period, it is possible to solve the NUM problem in a centralized manner and then deploy the solution to all the network nodes.
When some of the parameters which would affect the solution of the NUM problem changes, instead of solving the NUM problem again, a better way is to update the existing solution to approach the new optimal value of the NUM problem. This is also called a dynamic control algorithm.

Dynamic control algorithms are an important future research direction. 
The centralized solution of solving the NUM problem provides an upper bound on the performance of the dynamic algorithms. We hope the dynamic control algorithm has an equilibrium that is close to the optimal value of the NUM problem. Moreover, as we have seen from the literature, the algorithm for solving the NUM problem may provide guidances about the design of the control algorithm.

In another direction of future research, variations of our NUM problem can be studied by considering more communication features. In the existing research, multi-path communications, multicast communications, power control, routing, multi-radio, multi-channel and so on can all be incorporated in the NUM framework. Some of the combinations of the communication features are also valid and valuable for using BATS codes as well.

%
%
%

%
%

%
%

%
%


\newpage
\appendices

\section{Discussion of Batch-wise Packet Loss Model}
\label{app:lossmodel}

Consider an edge $e=(u,v)$ and a batch of rank $r$. Suppose $t$ packets of the batch are transmitted by node $u$ on edge $e$ using a certain linear recoding approach, where the order of these $t$ packets for transmission is uniformly at random decided as among all the permutations of $t$ elements.
Let $\tilde E_r(k)$ be the expected rank of the batch at node $v$ when $ k $ packets are received.

\begin{lemma}
$ \tilde E_r(k) $ is a non-decreasing, concave function of $ k $.
\end{lemma}
\begin{IEEEproof}
  Let $\Phi$ be the $r\times t$ recoding matrix for generating the $t$ recoded packets. Fix $k=0,1,\ldots,t-1$. Let $B$ be an $r\times (k+1)$ submatrix of $\Phi$ formed by $k+1$ columns chosen uniformly at random.
  Let $B'$ be the first $k$ columns of $B$ and $b$ be the last column of $B$.
  Then $ \tilde E_r(k) = \E [\rank (B')] $ and $ \tilde E_r(k+1) =\E [\rank (B)]$.
  Denote by $ \col(A) $ the column space of a matrix $ A $. We have
  \begin{IEEEeqnarray*}{rCl}
    \tilde E_r(k+1))- \tilde E_r(k) & =& \E(\rank([B'\ b]))-\E(\rank(B'))\\
                                    &=&\E(\rank([B'\ b])-\rank(B'))\\
                                    &=& \Pr\{b\notin \col(B')  \}\\
                                    &\geq& 0,
  \end{IEEEeqnarray*}
  i.e., $\tilde E_r(k)$ is non-decreasing. 

  Fix $k=0,1,\ldots,t-2$. Let $A$ be an $r\times (k+2)$ submatrix of $\Phi$ formed by $k+2$ columns chosen uniformly at random.
  Let $A'$ be the first $k$ columns of $A$, $a$ be the second last column of $A$ and $b$ be the last column of $A$.
  Then $ \tilde E_r(k) = \E [\rank (A')] $, $ \tilde E_r(k+1) =\E [\rank ([A'\ a])]=\E [\rank ([A'\ b])]$, and $ \tilde E_r(k+2) =\E [\rank ([A'\ a\ b])]$.
Then,
	\begin{IEEEeqnarray*}{rCl}
         \IEEEeqnarraymulticol{3}{l}{( \tilde E_r(k+2)- \tilde E_r(k+1))-( \tilde E_r(k+1)- \tilde E_r(k))}\\
	&=& (\E( \rank([A'\ a\ b])) -\E( \rank([A'\ b])) -(\E( \rank([A'\ a]))-\E( \rank(A')) \\
	&=&\E(\rank([A'\ a\ b])-\rank([A'\ b])  )-\E(\rank([A'\ a])-\rank(A')  )\\
	&=& \Pr\{a\notin  \col([A'\ b])\}-\Pr\{a\notin  \col(A')\}\\
	&\leq& 0,
	\end{IEEEeqnarray*}
        i.e., $\tilde E_r(k)$ is concave.
\end{IEEEproof}

For the batch-wise packet loss mode $q_e$, let
\begin{equation*}
E_r(t) = \sum_{i=0}^t q_e(i|t) \tilde E_r(i),
\end{equation*}
i.e., $ E_r(t) $ be the expected rank of this batch at the next network node $ v $. 
When $ E_r(t) $ is a non-decreasing, concave function with respect to $ t $, an algorithm in~\cite{yin_adaptive19} for solving \eqref{eq:aropt} can guarantee to converge to the optimal solution, and the solution is almost deterministic. 
In~\cite{yin_adaptive19}, it has been argued that when the packet loss pattern is a stationary process, $ E_r(t) $ is a non-decreasing, concave function of $ t $ and hence the solution of \eqref{eq:aropt} is almost deterministic. Here we show a more general condition on the batch-wise packet loss model so that $E_r(t)$ is a non-decreasing, concave function of $ t $.
Let 
\begin{equation}\label{eq:f}
  F(i|t) = \sum_{j\geq i} q_e(j|t). 
\end{equation}

\begin{theorem}

	If there exist $A_{i,t}\geq 0$ and $B_{i,t}\geq 0$ such that $B_{t+1,t}=0$, $ A_{0,t}=0 $ and for $i=0,1,\ldots, t+1$,
	\begin{equation}\label{eq:a1}
	F(i+1|t+2) - F(i+1|t+1) =  A_{i,t} + B_{i,t},
	\end{equation}
	and for $i=1,\ldots,t+1$,
	\begin{equation}\label{eq:a2}
	F(i|t+1) - F(i|t) = A_{i,t}+B_{i-1,t},
	\end{equation}
	then $ E_r(t) $ is a non-decreasing, concave function of $ t $.
\end{theorem}
\begin{IEEEproof}
  Let $\tilde E_r'(i) = \tilde E_r(i+1) - \tilde E_r(i)$.
  We can write
  \begin{equation*}
    E_r(t) = \sum_{i=1}^t F(i|t) \tilde E_r'(i-1).
  \end{equation*}
  Hence,
  \begin{IEEEeqnarray*}{rCl}
    E_r(t+2) - E_r(t+1) & = & \sum_{i=1}^{t+2}[F(i|t+2)- F(i|t+1)] \tilde E_r'(i-1) \\
    & = & \sum_{i=0}^{t+1}[F(i+1|t+2)- F(i+1|t+1)] \tilde E_r'(i) \\
    & = & \sum_{i=1}^{t+1}A_{i,t} \tilde E_r'(i) + \sum_{i=0}^{t} B_{i,t} \tilde E_r'(i) \IEEEyesnumber \label{eq:a3} \\
    & = & \sum_{i=0}^{t}A_{i+1,t} \tilde E_r'(i+1) + \sum_{i=0}^{t} B_{i,t} \tilde E_r'(i) \\
    & \leq & \sum_{i=0}^{t}A_{i+1,t} \tilde E_r'(i) + \sum_{i=0}^{t} B_{i,t} \tilde E_r'(i) \IEEEyesnumber \label{eq:a4} \\
    & = & \sum_{i=0}^{t}[A_{i+1,t}+B_{i,t}] \tilde E_r'(i) \\
    & = & \sum_{i=1}^{t+1}[A_{i,t}+B_{i-1,t}] \tilde E_r'(i-1) \\
    & = & \sum_{i=1}^{t+1}[F(i|t+1)-F(i|t)] \tilde E_r'(i-1) \IEEEyesnumber \label{eq:a5} \\
    & = & E_r(t+1) - E_r(t).
  \end{IEEEeqnarray*}
  where \eqref{eq:a3} follows from \eqref{eq:a1}, \eqref{eq:a4} is obtained due to $\tilde E_r(i)$ is concave of $i$, and \eqref{eq:a5} follows from \eqref{eq:a2}. By \eqref{eq:a3}, we see $E_r(t)$ is non-decreasing as $\tilde E_r'(i)\geq 0$.
\end{IEEEproof}

Let $ \{Z_i \} $ be the stochastic process which characterizes the packets loss pattern of the edge $ e $, i.e., $ Z_i=1 $ if the packet is received and $ Z_i=0 $, otherwise. We have \eqref{eq:a1} and \eqref{eq:a2} hold when $ \{Z_i \} $ is stationary, which can be verified in the following:
\begin{IEEEeqnarray*}{rCl}
	F(i+1|t+2) - F(i+1|t+1) &=& \Pr \{Z_1+\cdots+Z_{t+2}\geq i+1 \} -  \Pr \{Z_1+\cdots+Z_{t+1}\geq i+1 \}\\
	&=& \Pr \{Z_1+\cdots+Z_{t+1}= i , Z_{t+2}=1 \} \\
	&=& \Pr \{Z_1=1, Z_2+\cdots+Z_{t+1}= i -1, Z_{t+2}=1 \} +\\
	&& \Pr \{Z_1=0, Z_2+\cdots+Z_{t+1}= i , Z_{t+2}=1 \} \\
	&=& A_{i,t}+B_{i,t},
\end{IEEEeqnarray*}
and 
\begin{IEEEeqnarray}{rCl}
	F(i|t+1)-F(i|t) &=& \Pr \{Z_1+\cdots+Z_{t+1}\geq i \} -  \Pr \{Z_1+\cdots+Z_{t}\geq i \}\nonumber\\
	&=& \Pr \{Z_1+\cdots+Z_{t}= i-1 , Z_{t+1}=1 \} \nonumber\\
	&=& \Pr \{Z_2+\cdots+Z_{t+1}= i-1 , Z_{t+2}=1 \} \nonumber\\
	&=& \Pr \{Z_1=1, Z_2+\cdots+Z_{t+1}= i -1, Z_{t+2}=1 \} +\nonumber\\
	&& \Pr \{Z_1=0, Z_2+\cdots+Z_{t+1}= i-1 , Z_{t+2}=1 \} \nonumber\\
	&=& A_{i,t}+B_{i-1,t},\nonumber
\end{IEEEeqnarray}
where 
\begin{IEEEeqnarray}{rCl}
	A_{i,t} &= & \Pr \{Z_1=1, Z_2+\cdots+Z_{t+1}= i -1, Z_{t+2}=1 \} , \\
	B_{i,t}&=&\Pr \{Z_1=0, Z_2+\cdots+Z_{t+1}= i , Z_{t+2}=1 \}.
      \end{IEEEeqnarray}

      \section{Properties of Expected Rank Function}
      \label{app:erf}

Consider a length-L line network with node set $ \mc V=\{v_0,\ldots, v_L\} $ and edge set $ \mc E=\{e_{\ell}=(v_{\ell-1},v_{\ell}),\ell=1,\ldots,L\} $. 
Suppose a batch is transmitted in this line network, for $ \ell=1,\ldots,L $, the recoding generator matrix of the node $ v_{\ell-1} $ is $ \Phi_{\ell} $ and $ E_{\ell}$ is a diagonal matrix, in which the diagonal elements represent the packet loss pattern of the transmitted packets in link $ e_\ell  $. Then the rank of this batch at the destination node is equal to $ \rank (\Phi_1 E_1\cdots \Phi_\ell E_\ell\cdots \Phi_{L}E_{L}) $. Suppose the recoding number at each link is $ \{m_{e_\ell}\} $, $ \ell=1,\ldots,L $. Let $ \bar{h}(m_{e_\ell}) $ be the expected rank of the batch at the destination node when the recoding number in $ v_{\ell-1} $ is $ m_{e_\ell} $ and the recoding numbers in other nodes are fixed to be $ m_{e_j} $ for $ j\neq \ell $.

	In the following lemma and theorem, uniformly random coding is employed in the recoding scheme of each node, which implies $ \{\Phi_\ell\}_{\ell=1}^L $ are all uniformly random matrices. We use $ F(i|t) $ as defined in \eqref{eq:f}.
\begin{theorem}
		If for any $  t\geq 0 $ and $ i=0,1,\ldots,t+1 $,
	\begin{equation*}
	F(i+1|t+2)-F(i+1|t+1)\geq 0,
	\end{equation*}
	$ \bar{h}(m_{e_\ell})  $ is a non-decreasing function of $ m_{e_\ell} $ when uniformly random coding is employed in the recoding scheme of each node.
\end{theorem}
\begin{IEEEproof}
	Let $ \tilde{E}_r^L(k) $ be the expected rank of the batch at the destination node $ v_L $ when the rank of the batch equals $ r $ at node $ v_{\ell-1} $ and $ k $ packets are received at node $ v_{\ell} $. 
	For the batch-wise packet loss mode $q_e$, let
	\begin{IEEEeqnarray*}{rCl}
		E^L_r(m_{e_\ell}) = \sum_{k=0}^{m_{e_\ell}} q_e(k|m_{e_\ell}) \tilde E^L_r(k),
	\end{IEEEeqnarray*}
	i.e., $ E^L_r(m_{e_\ell}) $ be the expected rank of this batch at the destination node $ v_L $ when the rank of the batch equals $ r $ at node $ v_{\ell-1} $ and the recoding number at node $ v_{\ell-1} $ is $ m_{e_\ell}$. Then 
	\begin{IEEEeqnarray*}{rCl}
		\bar{h}(m_{e_\ell})=  \sum_{r=0}^{M} h_{v_{\ell-1}} (r) E_r^L(m_{e_\ell}).
	\end{IEEEeqnarray*}

Firstly, we will show $ \tilde E^L_r(k) $ is a non-decreasing function of $ k $.
	Let $ H=\Phi_1 E_1 \cdots \Phi_{\ell-1} E_{\ell-1} $ and $ U=E_{\ell+1}\Phi_{\ell+2}E_{\ell+2}\cdots \Phi_{L}E_L $.
	Let $  \theta_\ell=\{j\mid j'\mathrm{th}\text{ column of $ E_\ell $ is zero}\}$. $ \Phi_\ell $ is a uniformly random matrix  of size $ m_{e_{\ell-1}}\times m_{e_\ell} $.
	\begin{IEEEeqnarray*}{rCl}
	\tilde{E}_r^L(k) &=& \E [\rank(H\Phi_\ell E_\ell \Phi_{\ell+1}U)\mid \rank(H)=r, |\theta_\ell|=m_{e_\ell}-k]\\
	&=&  \E [\rank(H^*\Phi_\ell^*E_\ell\Phi_{\ell+1}U)\mid |\theta_\ell|=m_{e_\ell}-k]\\
	&=&  \E [ \rank(\Phi_\ell^*E_\ell\Phi_{\ell+1}U) \mid |\theta_\ell|=m_{e_\ell}-k ],
	\end{IEEEeqnarray*}
	where $ H^* $ is formed by $ r $ linearly independent columns of $ H $ and $ \Phi_\ell^* $ is an $ r\times m_{e_\ell} $ matrix such that $ H \Phi_\ell =H^* \Phi_\ell^* $. Then $ \Phi_\ell^* $ is a uniformly random matrix \cite{yang17monograph}. 
	We rewrite 
	\begin{IEEEeqnarray*}{rCl}
		\tilde{E}_r^L(k) &=&\E [(\rank\big((\Phi_\ell^*E_\ell)'\Phi_{\ell+1}'U\big) \mid |\theta_\ell|=m_{e_\ell}-k ],
	\end{IEEEeqnarray*}
	where $ (\Phi_\ell^*E_\ell)' $ a submatrix of $ \Phi_\ell^*E_\ell $ which is obtained by taking out the columns in $ \Phi_\ell^*E_\ell $ with the column index belonging to $ \theta_\ell $ and $ \Phi_{\ell+1}' $ is a submatrix of $ \Phi_{\ell+1} $ which is given by putting off the rows in $ \Phi_{\ell+1} $ of row index in $ \theta_\ell $. $ \Phi_{\ell+1}' $ is also a uniformly random matrix of size $ k\times m_{e_{\ell+1}} $.
	
	 Let $ B$ be a $ r\times (k+1) $ submatrix of $ \Phi^*_\ell $ formed by $ k+1 $ columns chosen uniformly at random. $ B' $ is the first $ k $ columns of $ B $ and $ b $ is the last column of $ B. $
	$ \Phi_{\ell+1}$ is a uniformly random matrices of size $ (k+1)\times m_{e_{\ell+1}} $.
	 Then
	\begin{IEEEeqnarray*}{rCl}
		\tilde{E}_r^L(k+1) = \E \big[\rank\big((B',b) \Phi_{\ell+1}U\big)   \big],\quad
		\tilde{E}_r^L(k) = \E \big[\rank(B'\Phi'_{\ell+1}U)   \big].
	\end{IEEEeqnarray*} 
	\begin{IEEEeqnarray*}{rCl}
		&&\tilde{E}_r^L(k+1)-\tilde{E}_r^L(k) \\
		&=& \E \big[\rank\big((B',b) \Phi_{\ell+1}U\big) \big] - \E \big[\rank(B'\Phi'_{\ell+1}U )  \big] \\
		&=& \E \big[  \rank\big((B',b) \bigl( \begin{smallmatrix}
			{\Phi'}_{\ell+1}\\ \phi
		\end{smallmatrix} \bigr)
	U\big)  - \rank(B'\Phi'_{\ell+1}U )  \big] \\
	&=&   \Pr \{b\in \col(B')\}   \E \big[  \rank\big((B'\Phi_{\ell+1}'+b\phi)
	U\big)  - \rank(B'\Phi'_{\ell+1}U ) \mid b\in \col(B') \big] +
	\\
	&&\Pr \{b\not\in \col(B')\}   \E \big[  \rank\big((B'\Phi_{\ell+1}'+b\phi)
	U\big)  - \rank(B'\Phi'_{\ell+1}U ) \mid b\not\in \col(B') \big] \\
		   &\geq &0,
	\end{IEEEeqnarray*}
	where $ \phi $ is a uniformly random row vector. Note that the last inequality holds since when $ b\not\in \col (B')$,
	\begin{IEEEeqnarray*}{rCl}
	\rank(	(B'\Phi'_{\ell+1}+b\phi) U  )\geq \rank( B'\Phi'_{\ell+1}U   ),
	\end{IEEEeqnarray*}
	and when $ b\in \col(B') $, due to the fact that each column of $ B'\Phi_{\ell+1}'+b\phi $ and $ B'\Phi_{\ell+1}' $ is the vector chosen uniformly at random from $ \col(B') $ independent of $ U $, 
\begin{IEEEeqnarray*}{rCl}
	\rank(	(B'\Phi'_{\ell+1}+b\phi )U  )= \rank( B'\Phi'_{\ell+1}U   ).
\end{IEEEeqnarray*}

	Let ${ E_r'}^{L}(k) = \tilde E^L_r(k+1) - \tilde E^L_r(k)$.
	We can write
	\begin{equation*}
	E^L_r(m_{e_\ell}) = \sum_{k=1}^{m_{e_\ell}} F(k|m_{e_\ell}) { E_r'}^{L}(k-1).
	\end{equation*}
Then
	\begin{IEEEeqnarray*}{rCl}
		E^L_r(m_{e_\ell}+2) - E^L_r(m_{e_\ell}+1) & = & \sum_{k=1}^{m_{e_\ell}+2}[F(k|m_{e_\ell}+2)- F(k|m_{e_\ell}+1)]{ E_r'}^{L}(k-1) \\
		& = & \sum_{k=0}^{m_{e_\ell}+1}[F(k+1|m_{e_\ell}+2)- F(k+1|m_{e_\ell}+1)] { E_r'}^{L}(k) \\
		&\geq&0,
	\end{IEEEeqnarray*}
which implies $ E_r^L(m_{e_\ell}) $ is a non-decreasing function of $ m_{e_\ell} $ and hence $ \bar{h}(m_{e_\ell}) $ is non-decreasing.
\end{IEEEproof}

\section{Primal-Dual  Approach}\label{sec:primal_dual}

We discuss two primal-dual based algorithms for solving \eqref{eq:1}.

\subsection{The First Formula}

Associating a Lagrange multiplier $\lambda_e$ for each inequality constraint in \eqref{eq:1}, the Lagrangian is
\begin{IEEEeqnarray*}{C}
  L=\sum_i \left[ U_i(\alpha_i\E[\bh^i]) - \alpha_i \sum_{e\in \mc P^i} \lambda_e  \overline{m}_e^i \right]  + \sum_{e\in \mc E} \lambda_e s_e.
\end{IEEEeqnarray*}
Similar to \cite{lin2006tutorial}, we have the following iterative algorithm for the dual problem.
In iteration $t=1,2,\ldots$, for each flow $\mc P^i$, 
$(\alpha_i,  (p_e^i,e\in \mc P^i))$ is updated by
\begin{equation}\label{eq:update_recoding2}
  (\alpha_i, (p_e^i,e\in \mc P^i)) \leftarrow \argmax_{\alpha_i,  (p_e^i,e\in \mc P^i)} \left[U_i(\alpha_i\E[\bh^i])-\alpha_i \sum_{e\in \mc P^i} \lambda_e \overline{m}_e^i \right];
  \end{equation}
the scheduling rate vector $ s$ is updated by
  \begin{equation}\label{eq:update_scheduling2}
	  s \leftarrow \argmax_{s\in \cov(\mc R)}\sum_{e\in \mc E} \lambda_e s_e;
        \end{equation}
and the Lagrange multipliers are updated by
  \begin{equation}\label{eq:update_mul2}
	  \lambda_e \leftarrow \max\left\{0, \lambda_e + \gamma_t \left( \sum_{i\colon e\in \mc P^i}\alpha_i\overline{m}_e^i - s_e \right) \right\},
  \end{equation}
  where $\{\gamma_1, \gamma_2, \ldots\}$ is a sequence of positive step sizes such that the subgradient search converges, e.g., $ \sum_t \gamma_t = \infty $ and $ \sum_t \gamma_t^2 < \infty $.
 
The subproblem \eqref{eq:update_recoding2} is different from the traditional counterparts~\cite{lin2006tutorial,shakkottai2008network}.
Suppose $U_i$ is a logarithm function, we can further simplify \eqref{eq:update_recoding2} as follows:
\begin{IEEEeqnarray}{rCl}
  \IEEEeqnarraymulticol{3}{l}{\argmax_{\alpha_i, (p_e^i,e\in \mc P^i)} \left[U_i(\alpha_i\E[\bh^i])-\alpha_i \sum_{e\in \mc P^i} \lambda_e \overline{m}_e^i \right]} \nonumber\\
&=& \argmax_{(p_e^i,e\in \mc P^i)} \left\{\argmax_{\alpha_i}  \left[U_i(\alpha_i\E[\bh^i])-\alpha_i \sum_{e\in \mc P^i} \lambda_e \overline{m}_e^i \right] \right\}\nonumber\\
&=&  \argmax_{(p_e^i,e\in \mc P^i)} \left[U_i\left(\frac{\E[\bh^i]}{ \sum_{e\in \mc P^i} \lambda_e \overline{m}_e^i  }\right)\right] \label{eq:d0} \\
&=&  \argmax_{ (p_e^i,e\in \mc P^i)} \frac{\E[\bh^i]}{ \sum_{e\in \mc P^i} \lambda_e \overline{m}_e^i  } \label{eq:d1},
\end{IEEEeqnarray}
where \eqref{eq:d0} follows that $ U_i(\alpha_i\E[\bh^i])-\alpha_i \sum_{e\in \mc P^i} \lambda_e \overline{m}_e^i $ is concave in $\alpha_i$ and thus achieves the maximum when its partial derivative with respect to $\alpha_i$ equals $0$, i.e.,
\begin{IEEEeqnarray}{rCl}
    \frac{1}{\alpha_i}-\sum_{e\in \mc P^i} \lambda_e \overline{m}_e^i =0;\label{eq:d2}
\end{IEEEeqnarray}
and \eqref{eq:d1} follows that $U_i$ is increasing.
Therefore, we can solve \eqref{eq:update_recoding2} by first obtaining $p_e^i$ from \eqref{eq:d1} and then deriving $\alpha_i$ by substituting $p_e^i$ into \eqref{eq:d2}. In other words, we can decompose \eqref{eq:update_recoding2} as:
\begin{IEEEeqnarray}{rCl}
	(p_e^i,e\in \mc P^i)&\leftarrow&  \argmax_{ (p_e^i,e\in \mc P^i)} \frac{\E[\bh^i]}{ \sum_{e\in \mc P^i} \lambda_e \overline{m}_e^i  }, \label{eq:update_recoding_2}\\
	\alpha_i &\leftarrow& \frac{1}{\sum_{e\in \mc P^i} \lambda_e \overline{m}_e^i}.\label{eq:update_alpha_2}
\end{IEEEeqnarray}

Since the optimization subproblem \eqref{eq:update_recoding_2} is difficult to solve directly, we solve (AP) by a primal-dual approach, where $p_e^i$ is updated using gradient search.
Denote by $\mc I$ the collection of all $(M+1)\times (M_0+1)$ stochastic matrices where the $(r,m)$ entry  is $p(m|r)$ for $m=0,1,\ldots,M_0$ and $r=0,1,\ldots,M$ such that $p(0|0)=1$ and $\sum_{m}p(m|r)=1$. For an $(M+1)\times (M_0+1)$ matrix $A$, denote by $\mathrm{proj}_{\mc I}(A)$ the projection of $A$ onto $\mc I$, which is the point in $\mc I$ that is closest to $A$.
 We obtain the projection by solving the convex optimization problem:
$ \min_{B\in \mathcal{I} } \|A-B\|^2_2 $ ($ \|\cdot \|_2 $ is the squared norm), which can be solved numerically. 
In iteration $t=1,2,\ldots$ of the primal-dual approach, $p_e^i$ is updated by
\begin{equation}
p_e^{i} \leftarrow \mathrm{proj}_{\mc I}\left(p_e^{i}+ \beta_t \frac{\partial}{\partial p_e^{i}}\frac{\E[\bh^i]}{ \sum_{e\in \mc P^i} \lambda_e \overline{m}_e^i  }\right), \label{eq:primal_recoding}
\end{equation}
where $\beta_t>0$ is the step size.
The updates of $\alpha_i$, $s$ and $\lambda_e$ are given in \eqref{eq:update_alpha_2}, \eqref{eq:update_scheduling2} and \eqref{eq:update_mul2}, respectively.
The $(r,m)$ entry of the gradient of $\E[\bh_{v_L}]$ with respect to $p_{e_l}$ is
\begin{IEEEeqnarray}{rCl}
	\frac{\partial \E[\bh_{v_L}] }{\partial p_{e_l}(m|r)} & = & \bh_{v_0} P_{v_1}\cdots P_{v_{l-1}} \frac{\partial P_{v_l}}{\partial p_{e_l}(m|r)} P_{v_{l+1}} \cdots  P_{v_L}[0,1,...,M]^T, \label{eq:gradient2}
\end{IEEEeqnarray}
where the $(i,j)$ entry of $\frac{\partial P_{v_l}}{\partial p_{e_l}(m|r)}$ is
\begin{IEEEeqnarray*}{rCl}
	\frac{\partial P_{v_l}[i,j]}{\partial p_{e_l}(m|r)}
	&=&
	\begin{cases}
		\sum_{k=j}^{m}q_{e_l}(k|m) \zeta^{i,k}_{j}& \text{if } i=r\text{ and }j\leq m, \\
		0 & \text{otherwise}.
	\end{cases}
\end{IEEEeqnarray*}

As the primal \eqref{eq:1} is nonconcave, the optimizer of the primal-dual approach may not be feasible for the primal one. Therefore, after obtaining a solution $(\tilde \alpha_i, (\tilde p_e^i,e\in \mc P^i))_i$ by multiple rounds of updating using the primal-dual approach,  we find a feasible primal solution by solving 
\eqref{eq:1} with $p_e^i=\tilde p_e^i$ fixed.

Moreover, motivated by the \emph{almost deterministic solutions} of adaptive recoding in~\cite{yin_adaptive19}, we may further require $p_e^i$ to be almost deterministic to reduce the complexity of the above approaches.

\subsection{Another Formula}

Put the stochastic constraints explicitly into \eqref{eq:1}
\begin{equation}
  \label{eq:1x}
\begin{IEEEeqnarraybox*}[][c]{rCl}
  \max_{(\alpha_i,  (p_e^i\geq 0,e\in \mc P^i))_i, s} & \quad & \sum_i U_i(\alpha_i\E[\bh^i]) \\
  \text{s.t.} & & \sum_{i\colon e\in \mc P^i} \alpha_i \overline{m}_e^i \leq s_e, \quad \forall e\in \mc E \\
  & & s=(s_e)\in \cov(\mc R) \\
  & & \sum_{m=1}^{M_0}p_e^i(m|r) \leq 1, \quad \forall i, e\in \mc P^i, r=1,\ldots,M.
\end{IEEEeqnarraybox*}
\end{equation}
The corresponding Lagrangian is
\begin{IEEEeqnarray*}{rCl}
  L & = & \sum_iU_i(\alpha_i\E[\bh^i]) - \sum_{e\in \mc E} \lambda_e\left(\sum_{i\colon e\in \mc P^i} \alpha_i \overline{m}_e^i - s_e \right)  - \sum_i\sum_{e\in \mc P^i} \theta_e^i ({p_e^i}' I_1-I_2)\\
  & = & \sum_i \left[ U_i(\alpha_i\E[\bh^i]) -  \sum_{e\in \mc P^i} \left(\lambda_e\alpha_i  \overline{m}_e^i + \theta_e^i ({p_e^i}' I_1-I_2)\right)\right]  + \sum_{e\in \mc E} \lambda_e s_e .
\end{IEEEeqnarray*}
where $I_1=(0,1,1,...,1)^T$ is a column vector of size $M_0+1$, $I_2$ is a size $M$ column vector with all elements being one and ${p_e^i}'$ is the submatrix formed by $1$st to $M$'th row of $p_e^i$.
Now, similar as \eqref{eq:d0}, we have
\begin{IEEEeqnarray}{rCl}
  \IEEEeqnarraymulticol{3}{l}{\argmax_{\alpha_i, (p_e^i\geq 0,e\in \mc P^i)} \left[U_i(\alpha_i\E[\bh^i])-\sum_{e\in \mc P^i} \left(\lambda_e\alpha_i  \overline{m}_e^i + \theta_e^i ({p_e^i}' I_1-I_2)\right)\right]} \nonumber\\
&=&  \argmax_{ (p_e^i\geq 0,e\in \mc P^i)} \left[U_i\left(\frac{\E[\bh^i]}{ \sum_{e\in \mc P^i} \lambda_e \overline{m}_e^i  }\right)-\sum_{e\in \mc P^i}\theta_e^i {p_e^i}' I_1\right]
\end{IEEEeqnarray}
As the above subsection, we use the primal-dual approach, which updates $p_e^i$ using gradient search. 
In each iteration, we update $p_e^i$ by
\begin{IEEEeqnarray}{rCl}
	p_e^{i} \leftarrow \left[p_e^{i}+ \beta_t \left(\frac{\frac{\partial}{\partial p_e^{i}}\frac{\E[\bh^i]}{ \sum_{e\in \mc P^i} \lambda_e \overline{m}_e^i  }}{ \frac{\E[\bh^i]}{ \sum_{e\in \mc P^i} \lambda_e \overline{m}_e^i  }  } -\frac{\partial }{\partial p_e^i}\theta_e^i {p_e^i}' I_1   \right)\right]^+,\nonumber\\
	&&\label{eq:primal_recoding_2}
\end{IEEEeqnarray}
where the submatrix formed by $1$st to $M$'th row of $\frac{\partial }{\partial p_e^i}\theta_e^i {p_e^i}' I_1$ equals ${\theta_e^i}^TI_1^T$ and $0$'th row equals zero.
The updates of $\alpha_i$, $\vv r$ and $\lambda_e$ still follow \eqref{eq:update_alpha_2}, \eqref{eq:update_scheduling2} and \eqref{eq:update_mul2}, respectively. Moreover, the update of $\theta_e^i$ is 
\begin{IEEEeqnarray}{rCl}
		\theta_e^i \leftarrow \left[\theta^i_e + \gamma_t \left(  p_e^i I_1-I_1  \right) \label{eq:primal_theta}\right]^+,
\end{IEEEeqnarray}
  where $\gamma_t$, $t=1,2,\ldots$ is a sequence of positive step sizes such that the subgradient search converges, e.g., $ \sum_t \gamma_t = \infty $ and $ \sum_t \gamma_t^2 < \infty $.
  
As the primal \eqref{eq:1} is nonconcave, the optimizer of the primal-dual approach may not be feasible for the primal one. Therefore, after obtaining a solution $(\tilde \alpha_i,  (\tilde p_e^i,e\in \mc P^i))_i$ by multiple rounds of updating using the primal-dual approach, we project $p_e^i$ to $\mc I$ to make it feasible.
Then we find a primal solution by solving 
\eqref{eq:1} with feasible $p^i_e$ fixed.


\begin{thebibliography}{10}
\providecommand{\url}[1]{#1}
\csname url@samestyle\endcsname
\providecommand{\newblock}{\relax}
\providecommand{\bibinfo}[2]{#2}
\providecommand{\BIBentrySTDinterwordspacing}{\spaceskip=0pt\relax}
\providecommand{\BIBentryALTinterwordstretchfactor}{4}
\providecommand{\BIBentryALTinterwordspacing}{\spaceskip=\fontdimen2\font plus
\BIBentryALTinterwordstretchfactor\fontdimen3\font minus
  \fontdimen4\font\relax}
\providecommand{\BIBforeignlanguage}[2]{{%
\expandafter\ifx\csname l@#1\endcsname\relax
\typeout{** WARNING: IEEEtran.bst: No hyphenation pattern has been}%
\typeout{** loaded for the language `#1'. Using the pattern for}%
\typeout{** the default language instead.}%
\else
\language=\csname l@#1\endcsname
\fi
#2}}
\providecommand{\BIBdecl}{\relax}
\BIBdecl

\bibitem{kelly1998rate}
F.~P. Kelly, A.~K. Maulloo, and D.~K.~H. Tan, ``Rate control for communication
  networks: shadow prices, proportional fairness and stability,'' \emph{Journal
  of the Operational Research society}, vol.~49, no.~3, pp. 237--252, 1998.

\bibitem{low1999optimization}
S.~H. Low and D.~E. Lapsley, ``Optimization flow control -- {I}: Basic
  algorithm and convergence,'' \emph{IEEE/ACM Transactions on networking},
  vol.~7, no.~6, pp. 861--874, 1999.

\bibitem{lin2006tutorial}
X.~Lin, N.~B. Shroff, and R.~Srikant, ``A tutorial on cross-layer optimization
  in wireless networks,'' \emph{IEEE J. Sel. Areas Commun.}, vol.~24, no.~8,
  pp. 1452--1463, 2006.

\bibitem{neill2008wnum}
D.~O'Neill, A.~Goldsmith, and S.~Boyd, ``Wireless network utility
  maximization,'' in \emph{MILCOM 2008 - 2008 IEEE Military Communications
  Conference}, 2008, pp. 1--8.

\bibitem{goldsmith1998amc}
A.~Goldsmith and S.-G. Chua, ``Adaptive coded modulation for fading channels,''
  \emph{IEEE Transactions on Communications}, vol.~46, no.~5, pp. 595--602,
  1998.

\bibitem{qiu99amc}
X.~Qiu and K.~Chawla, ``On the performance of adaptive modulation in cellular
  systems,'' \emph{IEEE Transactions on Communications}, vol.~47, no.~6, pp.
  884--895, 1999.

\bibitem{weber2005transmission}
S.~P. Weber, X.~Yang, J.~G. Andrews, and G.~De~Veciana, ``Transmission capacity
  of wireless ad hoc networks with outage constraints,'' \emph{IEEE
  transactions on information theory}, vol.~51, no.~12, pp. 4091--4102, 2005.

\bibitem{gao2009cross}
Q.~Gao, J.~Zhang, and S.~V. Hanly, ``Cross-layer rate control in wireless
  networks with lossy links: leaky-pipe flow, effective network utility
  maximization and hop-by-hop algorithms,'' \emph{IEEE Trans. Wireless
  Commun.}, vol.~8, no.~6, pp. 3068--3076, 2009.

\bibitem{flow}
R.~Ahlswede, N.~Cai, S.-Y.~R. Li, and R.~W. Yeung, ``Network information
  flow,'' \emph{{IEEE} Trans. Inform. Theory}, vol.~46, no.~4, pp. 1204--1216,
  Jul. 2000.

\bibitem{linear}
S.-Y.~R. Li, R.~W. Yeung, and N.~Cai, ``Linear network coding,'' \emph{{IEEE}
  Trans. Inform. Theory}, vol.~49, no.~2, pp. 371--381, Feb. 2003.

\bibitem{alg}
R.~Koetter and M.~Medard, ``An algebraic approach to network coding,''
  \emph{{IEEE/ACM} Trans. Networking}, vol.~11, no.~5, pp. 782--795, Oct. 2003.

\bibitem{random}
T.~Ho, B.~Leong, M.~Medard, R.~Koetter, Y.~Chang, and M.~Effros, ``The benefits
  of coding over routing in a randomized setting,'' in \emph{Proc. {IEEE}
  {ISIT} '03}, Jun. 2003.

\bibitem{Dana2006}
A.~F. Dana, R.~Gowaikar, R.~Palanki, B.~Hassibi, and M.~Effros, ``Capacity of
  wireless erasure networks,'' \emph{{IEEE} Trans. Inform. Theory}, vol.~52,
  no.~3, pp. 789--804, 2006.

\bibitem{Lun2008}
D.~S. Lun, M.~M\'edard, R.~Koetter, and M.~Effros, ``On coding for reliable
  communication over packet networks,'' \emph{Physical Communication}, vol.~1,
  no.~1, pp. 3--20, 2008.

\bibitem{wu06jsac}
{Yunnan Wu} and {Sun-Yuan Kung}, ``Distributed utility maximization for network
  coding based multicasting: a shortest path approach,'' \emph{IEEE J. Sel.
  Areas Commun,}, vol.~24, no.~8, pp. 1475--1488, 2006.

\bibitem{chen2007rc}
L.~Chen, T.~Ho, S.~H. Low, M.~Chiang, and J.~C. Doyle, ``Optimization based
  rate control for multicast with network coding,'' in \emph{IEEE INFOCOM 2007
  - 26th IEEE International Conference on Computer Communications}, 2007, pp.
  1163--1171.

\bibitem{khreishah2008rc}
A.~Khreishah, C.-C. Wang, and N.~B. Shroff, ``Optimization based rate control
  for communication networks with inter-session network coding,'' in \emph{IEEE
  INFOCOM 2008 - The 27th Conference on Computer Communications}, 2008, pp.
  81--85.

\bibitem{zhang2009mp}
X.~Zhang and B.~Li, ``Optimized multipath network coding in lossy wireless
  networks,'' \emph{IEEE Journal on Selected Areas in Communications}, vol.~27,
  no.~5, pp. 622--634, 2009.

\bibitem{traskov12}
D.~{Traskov}, M.~{Heindlmaier}, M.~{Medard}, and R.~{Koetter}, ``Scheduling for
  network-coded multicast,'' \emph{IEEE/ACM Trans. Netw.}, vol.~20, no.~5, pp.
  1479--1488, 2012.

\bibitem{chou03}
P.~A. Chou, Y.~Wu, and K.~Jain, ``Practical network coding,'' in \emph{Proc.
  Allerton Conf. Comm., Control, and Computing}, Oct. 2003.

\bibitem{gkan05}
C.~Gkantsidis and P.~Rodriguez, ``Network coding for large scale content
  distribution,'' in \emph{Proc. {IEEE} {INFOCOM} '05}, 2005.

\bibitem{maym06}
P.~Maymounkov, N.~J.~A. Harvey, and D.~S. Lun, ``Methods for efficient network
  coding,'' in \emph{Proc. Allerton Conf. Comm., Control, and Computing}, Sep.
  2006.

\bibitem{bin_expander15}
\BIBentryALTinterwordspacing
B.~Tang, S.~Yang, Y.~Yin, B.~Ye, and S.~Lu, ``Expander chunked codes,''
  \emph{EURASIP Journal on Advances in Signal Processing}, vol. 2015, no.~1,
  pp. 1--13, 2015. [Online]. Available:
  \url{http://dx.doi.org/10.1186/s13634-015-0297-8}
\BIBentrySTDinterwordspacing

\bibitem{Silva2009}
D.~Silva, W.~Zeng, and F.~R. Kschischang, ``Sparse network coding with
  overlapping classes,'' in \emph{Proc. NetCod '09}, Jun. 2009, pp. 74--79.

\bibitem{Heidarzadeh2010}
A.~Heidarzadeh and A.~H. Banihashemi, ``Overlapped chunked network coding,'' in
  \emph{Proc. ITW '10}, Jan. 2010, pp. 1--5.

\bibitem{yaoli11}
Y.~Li, E.~Soljanin, and P.~Spasojevic, ``Effects of the generation size and
  overlap on throughput and complexity in randomized linear network coding,''
  \emph{IEEE Trans. Inform. Theory}, vol.~57, no.~2, pp. 1111--1123, Feb. 2011.

\bibitem{yang11ac}
S.~Yang and R.~W. Yeung, ``Coding for a network coded fountain,'' in
  \emph{Information Theory Proceedings (ISIT), 2011 IEEE International
  Symposium on}, Saint Petersburg, Russia, July 31 - Aug. 5 2011, pp.
  2647--2651.

\bibitem{Mahdaviani12}
K.~Mahdaviani, M.~Ardakani, H.~Bagheri, and C.~Tellambura, ``Gamma codes: A
  low-overhead linear-complexity network coding solution,'' in \emph{Proc.
  NetCod '12}, Jun. 2012, pp. 125--130.

\bibitem{Mahdaviani13}
K.~Mahdaviani, R.~Yazdani, and M.~Ardakani, ``Overhead-optimized gamma network
  codes,'' in \emph{Proc. {NetCod} '13}, 2013.

\bibitem{bin_ldpc16}
B.~Tang and S.~Yang, ``An improved design of overlapped chunked codes,'' in
  \emph{Communications Proceedings (ICC), 2016 IEEE International Conference
  on}, 23-27, May 2016, pp. 1--6.

\bibitem{yang14bats}
S.~Yang and R.~W. Yeung, ``Batched sparse codes,'' \emph{{IEEE} Trans. Inform.
  Theory}, vol.~60, no.~9, pp. 5322--5346, Sep. 2014.

\bibitem{yang17monograph}
------, \emph{{BATS} Codes: Theory and Practice}, ser. Synthesis Lectures on
  Communication Networks.\hskip 1em plus 0.5em minus 0.4em\relax Morgan \&
  Claypool Publishers, 2017.

\bibitem{huang14mobihoc}
Q.~Huang, K.~Sun, X.~Li, and D.~O. Wu, ``Just {FUN}: A joint fountain coding
  and network coding approach to loss-tolerant information spreading,'' in
  \emph{Proc. of the 15th ACM Int. Symp. on Mobile Ad Hoc Netw. and Computing},
  2014, pp. 83--92.

\bibitem{yang14a}
S.~Yang, R.~W. Yeung, H.~F. Cheung, and H.~H.~F. Yin, ``{BATS}: Network coding
  in action,'' in \emph{2014 52nd Annual Allerton Conf. Commun., Control, and
  Computing}, Oct. 2014, pp. 1204--1211.

\bibitem{xu2016two}
X.~Xu, M.~S. G.~P. Kumar, Y.~L. Guan, and P.~H.~J. Chong, ``Two-phase
  cooperative broadcasting based on batched network code,'' \emph{IEEE Trans.
  Commun.}, vol.~64, no.~2, pp. 706--714, Feb. 2016.

\bibitem{yang18wuwnet}
S.~Yang, J.~Ma, and X.~Huang, ``Multi-hop underwater acoustic networks based on
  {BATS} codes,'' in \emph{Proc. of the 13th {ACM} Int. Conf. on Underwater
  Netw. {\&} Systems}, Dec. 2018, pp. 1--5.

\bibitem{yin20entropy}
H.~H.~F. Yin, R.~W. Yeung, and S.~Yang, ``A protocol design paradigm for
  batched sparse codes,'' \emph{Entropy}, vol.~22, no.~7, p. 790, 2020.

\bibitem{zhou19}
Z.~{Zhou}, C.~{Li}, S.~{Yang}, and X.~{Guang}, ``Practical inner codes for
  {BATS} codes in multi-hop wireless networks,'' \emph{IEEE Trans. Veh.
  Technol.}, vol.~68, no.~3, pp. 2751--2762, Mar. 2019.

\bibitem{zhou20}
Z.~Zhou, J.~Kang, and L.~Zhou, ``Joint {BATS} code and periodic scheduling in
  multihop wireless networks,'' \emph{IEEE Access}, vol.~8, pp.
  29\,690--29\,701, 2020.

\bibitem{tang16schedu}
B.~Tang, S.~Yang, B.~Ye, S.~Guo, and S.~Lu, ``Near-optimal one-sided scheduling
  for coded segmented network coding,'' \emph{IEEE Trans. Comput.}, vol.~65,
  no.~3, pp. 929--939, Mar. 2016.

\bibitem{yin_adaptive19}
H.~H.~F. Yin, B.~Tang, K.~H. Ng, S.~Yang, X.~Wang, and Q.~Zhou, ``A unified
  adaptive recoding framework for batched network coding,'' in \emph{Proc. IEEE
  ISIT '19}, Jul. 2019, pp. 1962--1966.

\bibitem{yin19overhearing}
H.~H.~F. Yin, X.~Xu, K.~H. Ng, Y.~L. Guan, and R.~W. Yeung, ``Packet efficiency
  of {BATS} coding on wireless relay network with overhearing,'' in \emph{Proc.
  IEEE ISIT '19}, Jul. 2019, pp. 1967--1971.

\bibitem{yin21intrablock}
H.~H.~F. Yin, K.~H. Ng, A.~Z. Zhong, R.~W. Yeung, and S.~Yang, ``Intrablock
  interleaving for batched network coding with blockwise adaptive recoding,''
  in \emph{Proc. IEEE ISIT '21}, Jul. 2021, pp. 1409--1414.

\bibitem{yin21impact}
H.~H.~F. Yin and K.~H. Ng, ``Impact of packet loss rate estimation on blockwise
  adaptive recoding for batched network coding,'' in \emph{Proc. IEEE ISIT
  '21}, Jul. 2021, pp. 1415--1420.

\bibitem{wang2021smallsample}
J.~Wang, Z.~Jia, H.~H.~F. Yin, and S.~Yang, ``Small-sample inferred adaptive
  recoding for batched network coding,'' in \emph{Proc. IEEE ISIT '21}, Jul.
  2021, pp. 1427--1432.

\bibitem{xu2018}
X.~Xu, Y.~L. Guan, and Y.~Zeng, ``Batched network coding with adaptive recoding
  for multi-hop erasure channels with memory,'' \emph{IEEE Trans. Commun.},
  vol.~66, no.~3, pp. 1042--1052, Mar. 2018.

\bibitem{yin_adaptive16}
H.~H.~F. Yin, S.~Yang, Q.~Zhou, and L.~M. Yung, ``Adaptive recoding for {BATS}
  codes,'' in \emph{Proc. IEEE ISIT '16}, Jul. 2016, pp. 2349--2353.

\bibitem{yin19recoding}
H.~F.~H. Yin, ``Recoding optimizations in batched sparse codes,'' Ph.D.
  dissertation, The Chinese University of Hong Kong, Jul. 2019.

\bibitem{gilbert1960capacity}
E.~N. Gilbert, ``Capacity of a burst-noise channel,'' \emph{Bell system
  technical journal}, vol.~39, no.~5, pp. 1253--1265, 1960.

\bibitem{elliott1963estimates}
E.~O. Elliott, ``Estimates of error rates for codes on burst-noise channels,''
  \emph{The Bell System Technical Journal}, vol.~42, no.~5, pp. 1977--1997,
  1963.

\bibitem{dong20icc}
Y.~Dong, S.~Jin, S.~Yang, and H.~H.~F. Yin, ``Network utility maximization for
  {BATS} code enabled multihop wireless networks,'' in \emph{Proc. {IEEE} {ICC}
  '20}, Jun. 2020.

\bibitem{Riley2010}
\BIBentryALTinterwordspacing
G.~F. Riley and T.~R. Henderson, \emph{The ns-3 Network Simulator}.\hskip 1em
  plus 0.5em minus 0.4em\relax Berlin, Heidelberg: Springer Berlin Heidelberg,
  2010, pp. 15--34. [Online]. Available:
  \url{https://doi.org/10.1007/978-3-642-12331-3_2}
\BIBentrySTDinterwordspacing

\bibitem{yang10bf}
S.~Yang, J.~Meng, and E.-h. Yang, ``Coding for linear operator channels over
  finite fields,'' in \emph{Proc. IEEE ISIT '10}, Jun. 2010, pp. 2413--2417.

\bibitem{kyasanur2006multichannel}
P.~Kyasanur, J.~So, C.~Chereddi, and N.~H. Vaidya, ``Multichannel mesh
  networks: challenges and protocols,'' \emph{IEEE Wireless Communications},
  vol.~13, no.~2, pp. 30--36, 2006.

\bibitem{shakkottai2008network}
S.~G. Shakkottai and R.~Srikant, \emph{Network optimization and control}.\hskip
  1em plus 0.5em minus 0.4em\relax Now Publishers Inc, 2008.

\bibitem{ephremides1990scheduling}
A.~Ephremides and T.~V. Truong, ``Scheduling broadcasts in multihop radio
  networks,'' \emph{IEEE Trans. Commun}, vol.~38, no.~4, pp. 456--460, April
  1990.

\bibitem{kelly1997charging}
F.~Kelly, ``Charging and rate control for elastic traffic,'' \emph{European
  transactions on Telecommunications}, vol.~8, no.~1, pp. 33--37, 1997.

\bibitem{low2017analytical}
S.~H. Low, \emph{Analytical methods for network congestion control}, ser.
  Synthesis Lectures on Communication Networks.\hskip 1em plus 0.5em minus
  0.4em\relax Morgan \& Claypool Publishers, 2017.

\bibitem{srikant2004mathematics}
R.~Srikant, \emph{The mathematics of Internet congestion control}.\hskip 1em
  plus 0.5em minus 0.4em\relax Springer Science \& Business Media, 2004.

\end{thebibliography}
\end{document}